\documentclass[11pt]{article}
\usepackage{times}
\usepackage{sidecap}
\usepackage{geometry}
\usepackage{floatrow}
\usepackage{float}
\usepackage{hyperref}
\usepackage{graphics,graphicx}
\usepackage{subcaption}
\usepackage{epsf,wrapfig}
\usepackage{stfloats}
\usepackage{utfsym}
\usepackage[bottom,para]{footmisc}

\usepackage{amssymb}

\usepackage{natbib,bm}
\usepackage{caption}
\usepackage{sidecap}
\sidecaptionvpos{figure}{c}

\DeclareGraphicsExtensions{.jpg,.pdf,.png,.eps}

\usepackage{caption}
\usepackage[dvipsnames]{xcolor}

\usepackage{tcolorbox}

\bibpunct{(}{)}{;}{a}{}{,}

\usepackage{soul}

\definecolor{darkcyan}{rgb}{0.0, 0.0, 0.7}

\definecolor{boxcolor}{RGB}{193, 229, 255}
\definecolor{arylideyellow}{rgb}{0.91, 0.84, 0.42}
\definecolor{applegreen}{rgb}{0.55, 0.71, 0.0}
\definecolor{asparagus}{rgb}{0.53, 0.66, 0.42}
\definecolor{babyblueeyes}{rgb}{0.63, 0.79, 0.95}
\definecolor{figboxcolor}{RGB}{193, 229, 255}
\definecolor{boxcolorgray}{RGB}{227, 227, 227}
\definecolor{darkcyan}{rgb}{0.0, 0.0, 0.7}

\sethlcolor{boxcolorgray}

\usepackage{paralist}
\let\olditem\item
\renewenvironment{thebibliography}[1]{%
  \section*{\refname}
  \let\par\relax
  \renewcommand{\item}[1][]{\olditem[\textbullet]}%
  \inparaenum}{\endinparaenum}

\newcounter{aff}

\DeclareGraphicsExtensions{.jpg,.pdf,.png,.eps}

\usepackage{caption}
\usepackage{xcolor}

\usepackage{tcolorbox}

\bibpunct{(}{)}{;}{a}{}{,}

\usepackage{soul}

\definecolor{darkcyan}{rgb}{0.0, 0.0, 0.7}

\definecolor{boxcolor}{RGB}{193, 229, 255}
\definecolor{figboxcolor}{RGB}{193, 229, 255}
\definecolor{boxcolorgray}{RGB}{227, 227, 227}
\definecolor{darkcyan}{rgb}{0.0, 0.0, 0.7}

\sethlcolor{boxcolorgray}

\def\gtrsim{\lower 2pt \hbox{$\, \buildrel {\scriptstyle >}\over
{\scriptstyle \sim}\,$}}
\def\lesssim{\lower 2pt \hbox{$\, \buildrel {\scriptstyle <}\over
{\scriptstyle \sim}\,$}}

\def\approxlt{\lower.2em\hbox{$\buildrel < \over \sim$}}
\def\approxgt{\lower.2em\hbox{$\buildrel > \over \sim$}}

\def\be{\begin{equation}}
\def\ee{\end{equation}}
\def\bea{\begin{eqnarray*}}
\def\eea{\end{eqnarray*}}

\def\jref@jnl#1{{\rm#1}}
\def\aj{\jref@jnl{AJ}}			
\def\araa{\jref@jnl{ARA\&A}}		
\def\apj{\jref@jnl{ApJ}}			
\def\apjl{\jref@jnl{ApJ}}		
\def\apjs{\jref@jnl{ApJS}}		
\def\ao{\jref@jnl{Appl.~Opt.}}		
\def\apss{\jref@jnl{Ap\&SS}}		
\def\aap{\jref@jnl{A\&A}}		
\def\aapr{\jref@jnl{A\&A~Rev.}}		
\def\aaps{\jref@jnl{A\&AS}}		
\def\azh{\jref@jnl{AZh}}			
\def\baas{\jref@jnl{BAAS}}		
\def\jrasc{\jref@jnl{JRASC}}		
\def\memras{\jref@jnl{MmRAS}}		
\def\mnras{\jref@jnl{MNRAS}}		
\def\pra{\jref@jnl{Phys.~Rev.~A}}	
\def\prb{\jref@jnl{Phys.~Rev.~B}}	
\def\prc{\jref@jnl{Phys.~Rev.~C}}	
\def\prd{\jref@jnl{Phys.~Rev.~D}}	
\def\pre{\jref@jnl{Phys.~Rev.~E}}	
\def\prl{\jref@jnl{Phys.~Rev.~Lett.}}	
\def\pasp{\jref@jnl{PASP}}		
\def\pasj{\jref@jnl{PASJ}}		
\def\qjras{\jref@jnl{QJRAS}}		
\def\skytel{\jref@jnl{S\&T}}		
\def\solphys{\jref@jnl{Sol.~Phys.}}	
\def\sovast{\jref@jnl{Soviet~Ast.}}	
\def\ssr{\jref@jnl{Space~Sci.~Rev.}}	
\def\zap{\jref@jnl{ZAp}}			
\def\nat{\jref@jnl{Nature}}		
\def\iaucirc{\jref@jnl{IAU~Circ.}}
\def\aplett{\jref@jnl{Astrophys.~Lett.}}
\def\apspr{\jref@jnl{Astrophys.~Space~Phys.~Res.}}
\def\bain{\jref@jnl{Bull.~Astron.~Inst.~Netherlands}}
\def\fcp{\jref@jnl{Fund.~Cosmic~Phys.}}
\def\gca{\jref@jnl{Geochim.~Cosmochim.~Acta}}
\def\grl{\jref@jnl{Geophys.~Res.~Lett.}}
\def\jcp{\jref@jnl{J.~Chem.~Phys.}}	
\def\jgr{\jref@jnl{J.~Geophys.~Res.}}	
\def\jqsrt{\jref@jnl{J.~Quant.~Spec.~Radiat.~Transf.}}
\def\memsai{\jref@jnl{Mem.~Soc.~Astron.~Italiana}}
\def\nphysa{\jref@jnl{Nucl.~Phys.~A}}
\def\physrep{\jref@jnl{Phys.~Rep.}}
\def\physscr{\jref@jnl{Phys.~Scr}}
\def\planss{\jref@jnl{Planet.~Space~Sci.}}
\def\procspie{\jref@jnl{Proc.~SPIE}}

\def\ion#1#2{#1$\;${\small\rm\@Roman{#2}}\relax}

\newcount\lecurrentfam
\def\LaTeX{\lecurrentfam=\the\fam \leavevmode L\raise.42ex
\hbox{$\fam\lecurrentfam\scriptstyle\kern-.3em A$}\kern-.15em\TeX}
\def\sizrpt{
(\fontname\the\font): em=\the\fontdimen6\font, ex=\the\fontdimen5\font
\typeout{
(\fontname\the\font): em=\the\fontdimen6\font, ex=\the\fontdimen5\font
}}
\definecolor{dark}{rgb}{ 0.34, 0.09, 0.49}

\usepackage{fancyhdr,url}
\usepackage{transparent}
\usepackage{tablefootnote}
\usepackage{mathpazo}
\usepackage{tabularx}
\usepackage{booktabs}
\usepackage{longtable}
\usepackage{booktabs}
\usepackage{tabularx}
\usepackage{threeparttable}
\usepackage{enumitem}

\usepackage{mathpazo}

\begin{document}

\noindent

\noindent
\begin{raggedright}
{\huge\bfseries Frontiers of chemical complexity\\  in the interstellar medium:  \\[0.2em]
\LARGE \vspace{2mm} \textit{unveiling the cosmic origins of life}
}\end{raggedright} 
\noindent
\vspace{1em}

Víctor M. Rivilla$^1$,
Miguel Sanz-Novo$^2$,
Maite Beltrán$^{3}$,
Jes J{\o}rgensen$^{4}$,
Laura Colzi$^{1}$, 
Alain Baudry$^{5}$,
Maria Drozdovskaya$^{6}$,
Timea Csengeri$^{5}$,
Sébastien Maret$^{7}$,
Frederique Motte$^{7}$,
Maryvonne Gerin$^{8}$,
Sergio Martín$^{9}$,
Audrey Coutens$^{10}$,
Claudia Toci$^{11}$,
Katharine Johnston$^{13}$,
Pamela Klaassen$^{14}$,
Eleonora Bianchi$^{3}$,
Claudio Codella$^{3}$,
Linda Podio$^{3}$,
Giovanni Sabatini$^{3}$,
Marta de Simone$^{15}$,
Francesco Fontani$^{3}$,
Álvaro Sánchez-Monge$^{16}$,
Izaskun Jiménez-Serra$^{1}$,
Aurore Bacmann$^{7}$,
Gary Fuller$^{17}$,
Peter Schilke$^{18}$,
Michael Küffmeier$^{4}$,
Shaoshang Zeng$^{19}$,
Adam Ginsburg$^{20}$,
Brett McGuire$^{21}$,
Pablo García$^{22,23}$,
Nami Sakai$^{19}$,
Maxime Tanious$^{16}$,
Korash Assani$^{6}$,
Miguel A. Requena-Torres$^{24,25}$

\vspace{1em}
{
\footnotesize{
\noindent $^1$ Centro de Astrobiología (CAB), CSIC-INTA, Carretera de Ajalvir km 4, Torrejón de Ardoz, 28850 Madrid, Spain 
$^2$ Max Planck Instutute for Extraterrestrial Physics, MPE, Germany 
$^3$ INAF-Osservatorio Astrofisico di Arcetri, Italy 
$^4$ Niels Bohr Institute, Denmark 
$^5$ University of Bordeaux, France 
$^6$ DCBP, Universität Bern, Switzerland
$^7$ Univ. Grenoble Alpes, France 
$^8$ LUX, Observatoire de Paris, CNRS, France 
$^9$ ESO/JAO, Chile 
$^{10}$ Univ. Toulouse, CNRS, IRAP, France
$^{11}$ University of Sevilla, Spain 
$^{13}$ University of Lincoln, United Kingdom 
$^{14}$ UK Astronomy Technology Centre, United Kingdom 
$^{15}$ ESO, Germany 
$^{16}$ ICE-CSIC, Spain 
$^{17}$ Jodrell Bank Centre for Astrophysics, The University of Manchester, United Kingdom 
$^{18}$ University of Cologne, Germany 
$^{19}$ RIKEN Pioneering Research Institute, Japan
$^{20}$ Univ. Florida, USA
$^{21}$ Massachusetts Institute of Technology, USA
$^{22}$ Chinese Academy of Sciences South America Center for Astronomy, National Astronomical Observatories, CAS, China
$^{23}$ Instituto de Astronom\'ia, Universidad Cat\'olica del Norte, Chile
$^{24}$ University of Maryland, USA,
$^{25}$ Towson University, USA
}
}

\vspace{2em}



\vspace{-0.5cm}
\begin{tcolorbox}[boxsep=3pt,left=0pt,right=0pt,top=1pt,bottom=1pt,colframe=boxcolor,colback=boxcolor]
\justify
The interstellar medium (ISM) contains a rich molecular inventory that provides a unique window into the chemistry operating during star and planet formation and its connection to prebiotic chemistry. More than 350 interstellar molecules are currently known, including increasingly complex organic molecules and species containing elements beyond the second row of the periodic table. Yet the limits that interstellar chemical complexity can reach, its evolution from molecular clouds to planetary systems, and its contribution to the molecular feedstock of life remain poorly constrained. 
%
Addressing these questions requires sensitive, broadband observations capable of detecting multiple weak rotational transitions of molecular species, such as amino acids, nucleobases, sugars and lipids, towards star- and planet-forming environments. Current facilities, including ALMA, are limited by instantaneous spectral coverage, sensitivity (especially on extended spatial scales), and the trade-off between total bandwidth and integration time. The proposed ALMA2040 facility would overcome these limitations through at least a tenfold improvement in spectral-line sensitivity, high-fidelity imaging from sub-arcsecond to tens-of-arcsecond scales, and ultra-broad instantaneous frequency coverage from 35 to 950 GHz at high spectral resolution. These capabilities would allow systematic detections of species with molecular abundances with respect to H$_2$ below 10$^{-11}$, 
facilitating the identification of new complex and heavy-element-bearing species through multiple transitions, and providing complete molecular inventories across evolutionary stages.
ALMA2040 would thus make it possible to comprehensively investigate the emergence and evolution of chemical complexity, and to determine whether molecular precursors of prebiotic chemistry are widespread throughout the Galaxy.
\end{tcolorbox}


\newpage{}

\section{Scientific Context and Open Questions}

\vspace{-2mm}



\subsection{The molecular Universe}

\vspace{-2mm}

\begin{figure*}[h!]
\centering
\includegraphics[width=0.95\textwidth]{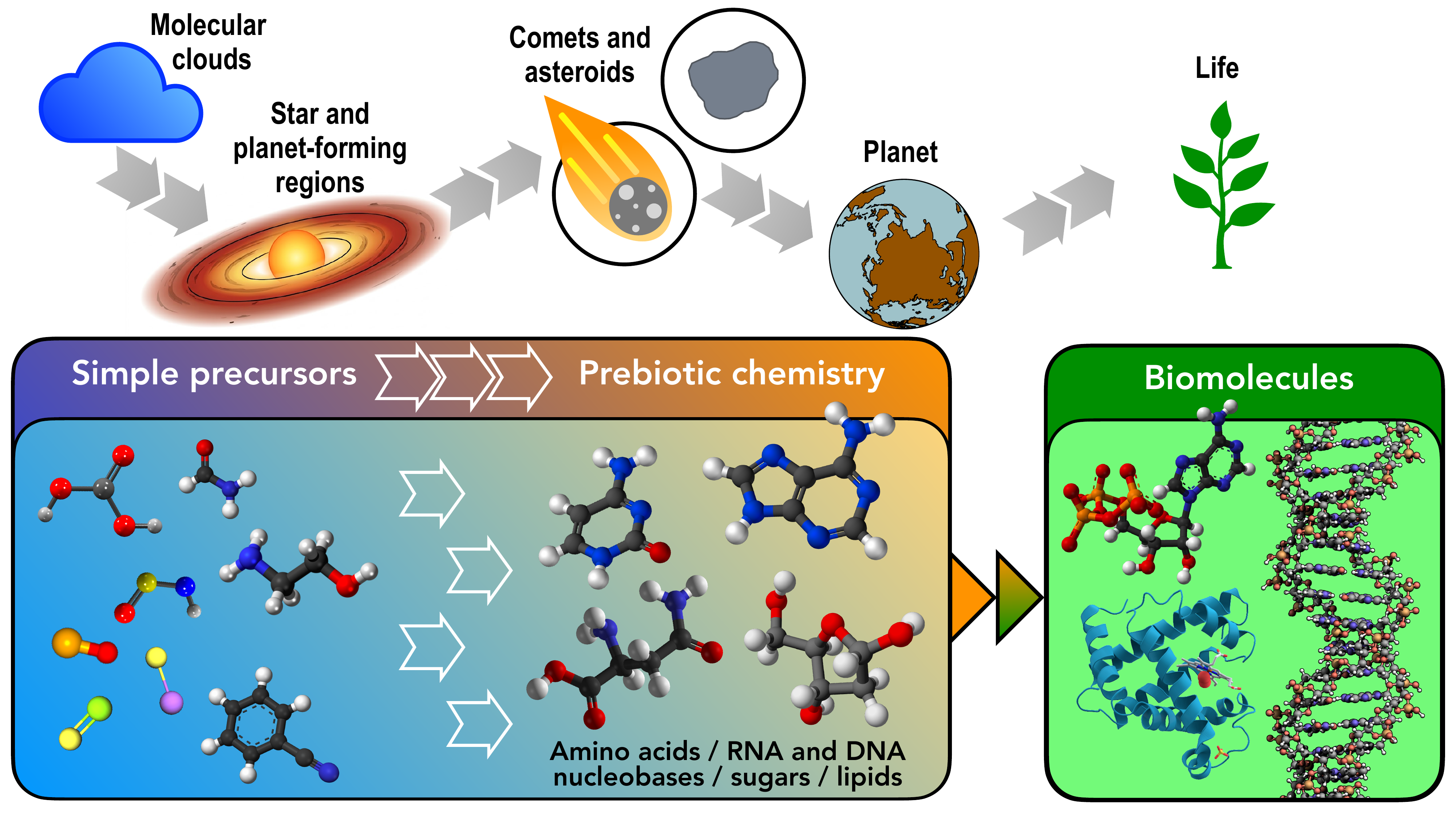}
\vspace{-5mm}
    \caption{\small From interstellar molecules to biomolecules.} 
    \label{fig:astro-to-bio} 
\end{figure*}

The first detection of an interstellar molecule, the methylidyne radical (CH), in the late 1930's (\citealt{dunham1937,swings1937}) 
started the time of {\it Molecular Astrophysics}. Nevertheless, until the end of the 1960s the number of interstellar species remained very low. Thanks to the advent of observational facilities with increasing sensitivity, mainly radiotelescopes, and the development of rotational spectroscopy in the laboratory, the pace of new detections took off. 
Around one third of the total number of molecules ($>$350\footnote{Updated lists are publicly available at \href{urlcolor=cyan}{https://cdms.astro.uni-koeln.de/classic/molecules}, and \href{urlcolor=cyan}{https://molecules-in.space/}.}) have been discovered in the last few years, including large species with tens of atoms (\citealt{wenzel2025}). 
%
%
%
%
The presence of molecules is ubiquitous in the Universe, from our Solar System to distant galaxies, and it pervades all the phases of the cycle of interstellar matter. These 
astronomical environments cover a broad range of evolutionary stages: atmosphere of (exo)planets\footnote{\citealt{madhusudhan2019}}, comets and asteroids\footnote{\citealt{bockelee-morvan1998,altwegg2015,Oba2023,glavin2025,furukawa2026,cordiner2026}}, diffuse/translucent clouds\footnote{\citealt{lucas1996,gerin2025}}, molecular clouds\footnote{\citealt{pratap1997,requena-torres2006,sakai2008,McGuire2020,Rivilla2021a,agundez2026}}, photo-dominated regions\footnote{\citealt{jansen1995,cuadrado2017}}, starless/prestellar/protostellar cores and clumps \footnote{\citealt{caselli1999,maret2005,fontani2011,Bacmann2012,belloche2013,vastel2014,jimenez-serra2016,jorgensen2016,rivilla2017,bianchi2019,colzi2022b}}, protoplanetary disks\footnote{\citealt{Dutrey2014,Booth2021,Brunken2022}}, circumstellar envelopes surrounding evolved stars\footnote{\citealt{cernicharo2000,ziurys2006}}, shocks\footnote{\citealt{bachiller1997,arce2008,Lefloch2017,matsuura2017,lopez-gallifa2025}}, and extragalactic environments\footnote{\citealt{martin2006,tercero2020,Sewilo2022,martin2021,yang2023}.}. 
%
%
This molecular reservoir offers the raw ingredients for the chemistry that will be developed during the formation of stars and planets (Fig. \ref{fig:astro-to-bio}). 

%

%
%


\vspace{-2mm}

\vspace{-8mm}
\subsection{The cosmic origins of life}
\vspace{-2mm}


Life appeared on Earth around 3.8 billion years ago, around 700 Myr after its formation (\citealt{pearce2018}), but the mechanisms that made it possible are still largely unknown. In particular, the processes that allowed the step from prebiotic chemistry to biology are uncertain. It is thought that some of the building blocks of prebiotic chemistry could have been endogenously synthetized on the planet through {\it in situ} mechanisms such as atmospheric reactions favored by ultraviolet radiation,  volcanic activity and/or hydrothermal vents in the oceans (e.g. \citealt{Benner2019}). Complementarily, some prebiotic material might have arrived to Earth through extraterrestrial delivery (\citealt{oro1961,Chyba&Sagan1992,cooper2001,pizzarello2006,rivilla2020}),
being thus available to contribute to prebiotic chemistry, and to the emergence of life (Fig. \ref{fig:astro-to-bio}).
Several independent lines of evidence support 
this exogenous contribution.
First, the detection of molecules directly relevant to biochemistry such as amino acids, nucleobases, sugars and lipids in comets and asteroids (e.g., \citealt{altwegg2015,glavin2025,furukawa2026}), which confirms that these basic organic compounds were present in our Solar System during the formation and growth of planets and planetesimals.
Second, 
%
%
%
laboratory experiments mimicking prebiotic conditions on early Earth have shown that pivotal biomolecules (e.g., the RNA monomers, ribonucleotides) can be synthesized starting from much smaller molecules (e.g., 
NH$_2$CONH$_2$, HOCH$_2$CN, NH$_2$OH, (CHOH)$_2$; \citealt{powner2009,patel2015,becker2019}).
Interestingly, many of these species have already been discovered in the ISM 
(e.g., \citealt{belloche2019,zeng2019,rivilla2020b,rivilla2022a}).
%
%
Therefore, {\bf a comprehensive understanding of the emergence and evolution of the interstellar molecular inventory across the different phases of star- and planet-forming regions 
is crucial to unveil the contribution that astrochemistry can have as a gateway to prebiotic chemistry}.  

\vspace{6mm}

\vspace{-10mm}
\subsection{The interstellar molecular zoo: what we know
so far?}
\vspace{-2mm}

\begin{wrapfigure}[17]{r}{0.675\textwidth}
    \centering
             \vspace{-4mm}
    \includegraphics[width=1\textwidth]{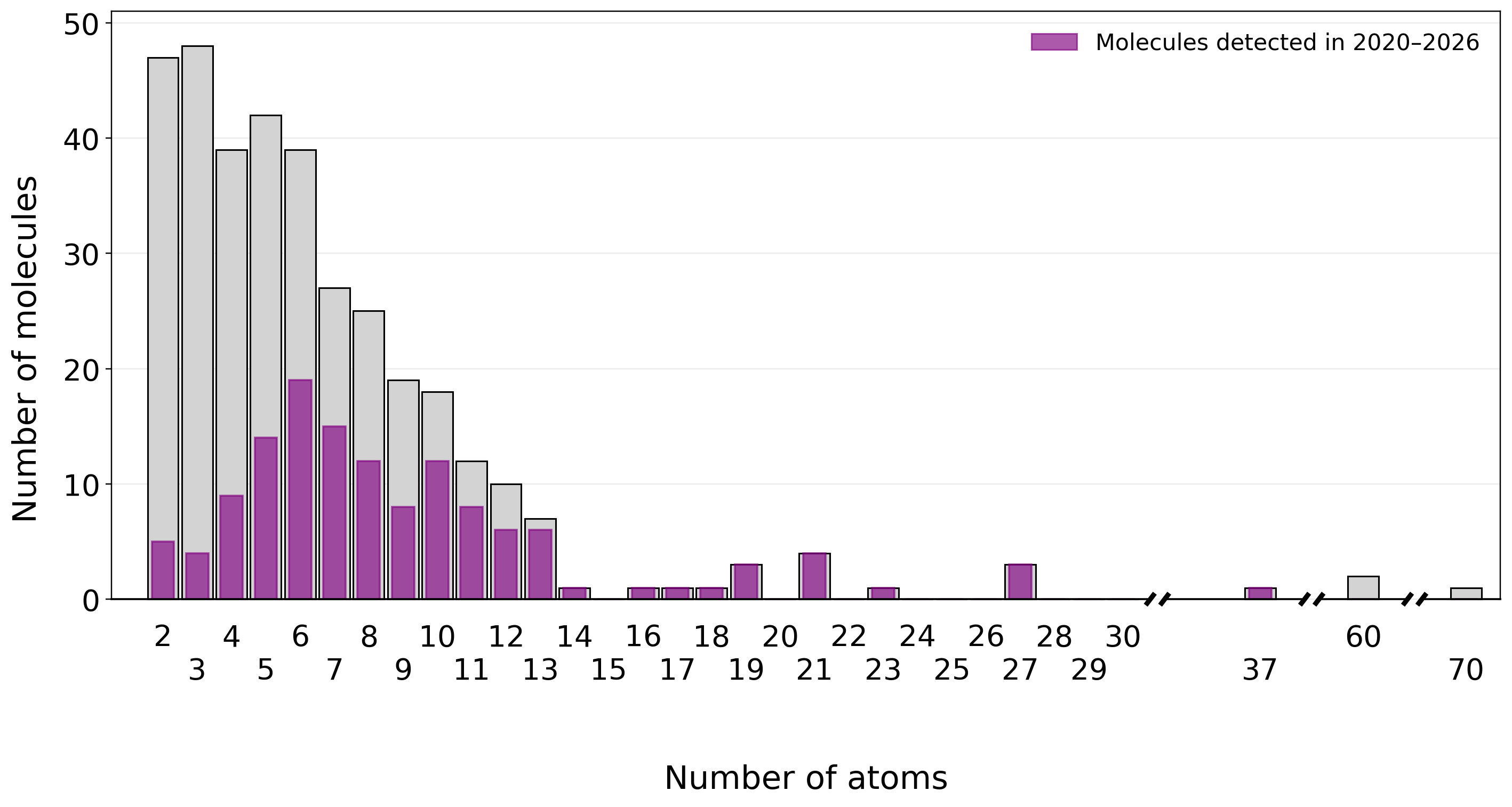}
        \vspace{-8mm}
    \caption{\small Distribution of interstellar molecules (as of October 2026) as a function of number of atoms.  
    Grey and purple bars represents the total census and those molecular species discovered only in the period 2020-2026, respectively.}
    \label{fig:interstellar-molecules}
\end{wrapfigure}

Nowadays, we are living a golden era for the discovery of new interstellar molecules  (\citealt{McGuire2022}). Fig. \ref{fig:interstellar-molecules} shows the number of molecules detected  in the ISM so far (as of October 2026), considering also circumstellar detections in the envelopes of evolved stars, as a function of the number of atoms. 
Strikingly, the most recent detections, between 2020 and 2026
are a high fraction of the total, especially for the largest species ($\geq$6 atoms, containing carbon), which were defined as {\it complex organic molecules (COMs)} by \citet{herbst2009}.
These recent discoveries are continuously pushing our knowledge about the size of the molecules that interstellar chemistry is able to synthesize. Large species of several chemical families have been reported: alcohols (\citealt{mcguire2017,Belloche2022,jimenez-serra2022,Fried2024}), amides (\citealt{rivilla2023}), imines (\citealt{bizzocchi2020}), isocyanates (\citealt{rodriguez-almeida2021a}), aldehydes (\citealt{coutens2022,sanz-novo2026}), nitriles (\citealt{belloche2014,zeng2019,Agundez2024}), or carboxylic acids (\citealt{sanz-novo2023}). 
%
In addition, species containing from one up to seven carbon rings, including polycyclic aromatic hydrocarbons (PAHs) and derivatives, have been detected (\citealt{mcguire2018_benzo,cernicharo2021_indene,cernicharo2024,wenzel2025,araki2026}).
%
%
%
%
However, as shown in Fig. \ref{fig:interstellar-molecules}, new smaller species have also been reported, such as HONO (\citealt{coutens2019}), HNSO (\citealt{Sanz-Novo2024b}), and several cations: PO$^+$ (\citealt{rivilla2022b}), CH$_3^+$, (\citealt{berne2023}), HOCS$^+$ (\citealt{sanz-novo2024}), and CCH$^+$ (\citealt{jacob2026}).
%

%

\begin{figure*}
\centering
\includegraphics[width=1.0\textwidth]{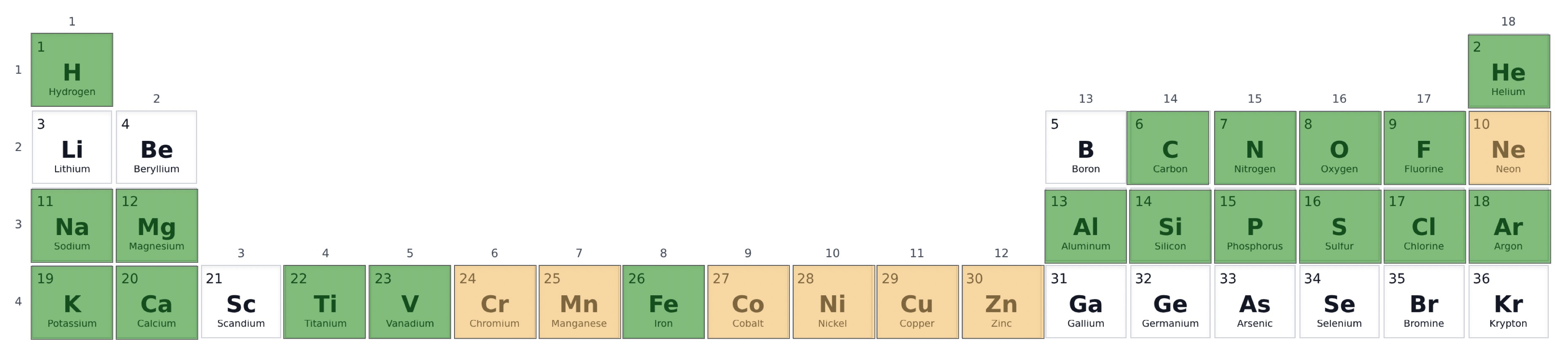}
\vspace{-10mm}
    \caption{\small {\it Astromolecular} table of chemical elements. The 19 elements highlighted in green 
    are those present in the molecules detected so far in the ISM (including circumstellar material). The 7 additional elements indicated in orange are among the 
    most cosmically abundant that have not yet been identified as constituents of interstellar molecules.} 
    \label{fig:astromolecular-table} 
\end{figure*}
\begin{wrapfigure}[17]{l}{0.6\textwidth}
    \centering
    \includegraphics[width=1\textwidth]{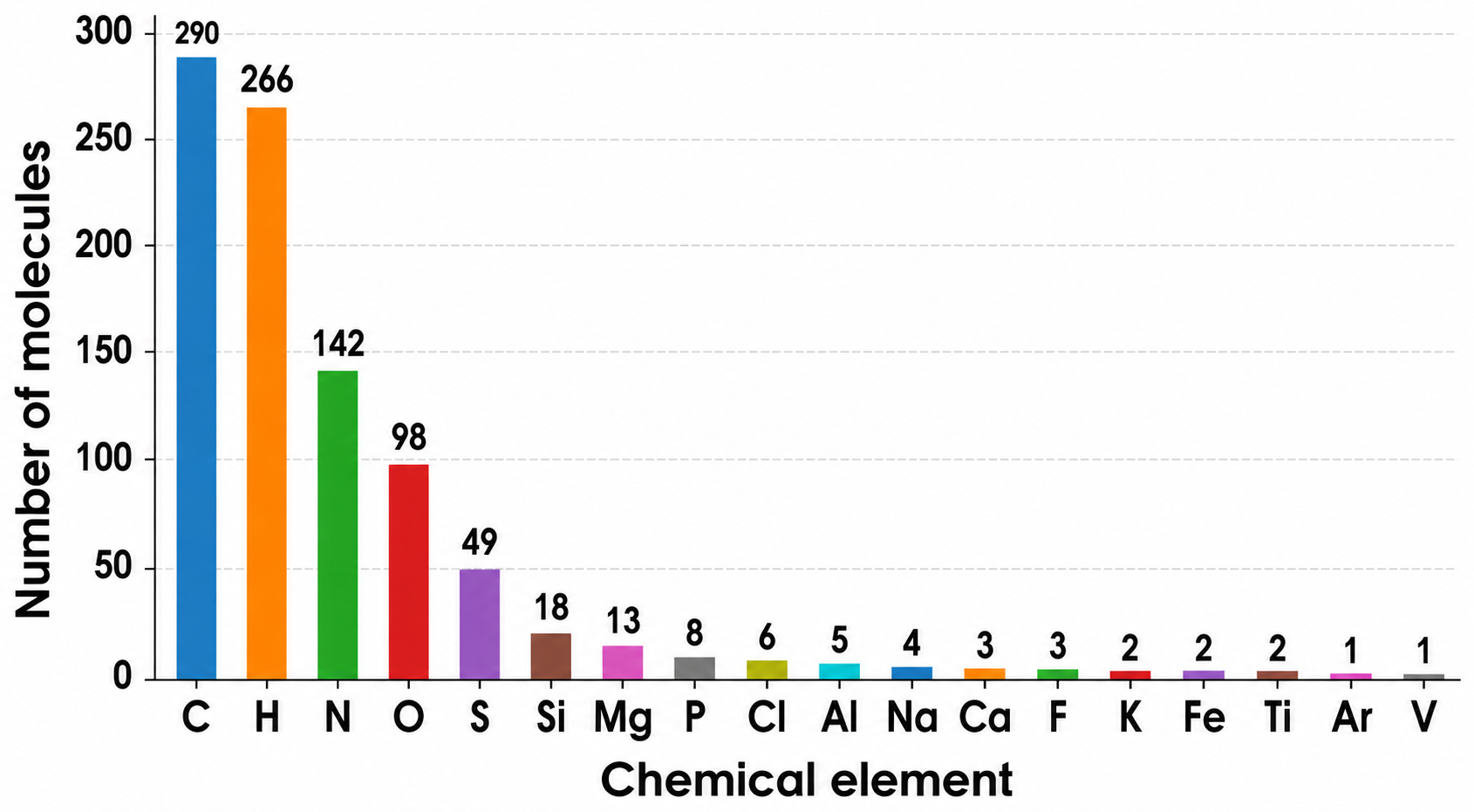}
    \vspace{-8mm}
    \caption{\small Number of molecules detected in the ISM (including also circumstellar material) containing a certain chemical element, ordered from highest to lower occurrence.}
    \label{fig:histo-elements}
\end{wrapfigure}

The definition of chemical {\it complexity} can be extended beyond its common meaning, which merely relies on the number of atoms, including also the  
variety of elemental constituents. Fig. \ref{fig:astromolecular-table} shows the {\it Astromolecular table of elements}, which depicts the 19 chemical elements that have been identified as constituents of interstellar species.
%
The list contains the 14 most cosmically abundant elements (\citealt{lodders2003}), including the six basic elements 
for life: carbon (C), hydrogen (H), nitrogen (N), oxygen (O), phosphorus (P) and sulfur (S), known as CHNOPS.
%
As shown in Fig. \ref{fig:histo-elements}, the most common elements in interstellar molecules are
%
several members of CHNOPS (C, H, N, O and S), while P is placed in the 8$^{th}$ position, after Si and Mg. This reinforces the link between the availability of chemical elements in the ISM and its relevance in prebiotic chemistry.

Although molecules containing elements in the third and fourth row of the periodic table such as Mg, Al, Na, K, Ca, Fe were first detected towards the very dense and hot circumstellar material surrounding evolved stars (e.g., \citealt{ziurys2006}), in recent years they have also been reported in regions affected by shocks, such as Galactic Center molecular clouds (NaS and MgS; \citealt{Rey-Montejo2024}), and hot disks around massive protostars (KCl and NaCl; \citealt{ginsburg2019,tanaka2020,zhang2026}; AlO, TiO and CaS, \citealt{tachibana2019,zhang2026,Tasa-Chaveli2025}). 
These heavier elements are thought to be released from refractory dust material through thermal evaporation in hot regions and/or shock-induced processes such as sputtering and grain disruption, and subsequently incorporated into gas-phase molecules.
%

\begin{wrapfigure}[22]{r}{0.53\textwidth}
    \centering
    \vspace{-6mm}
    \includegraphics[width=1\textwidth]{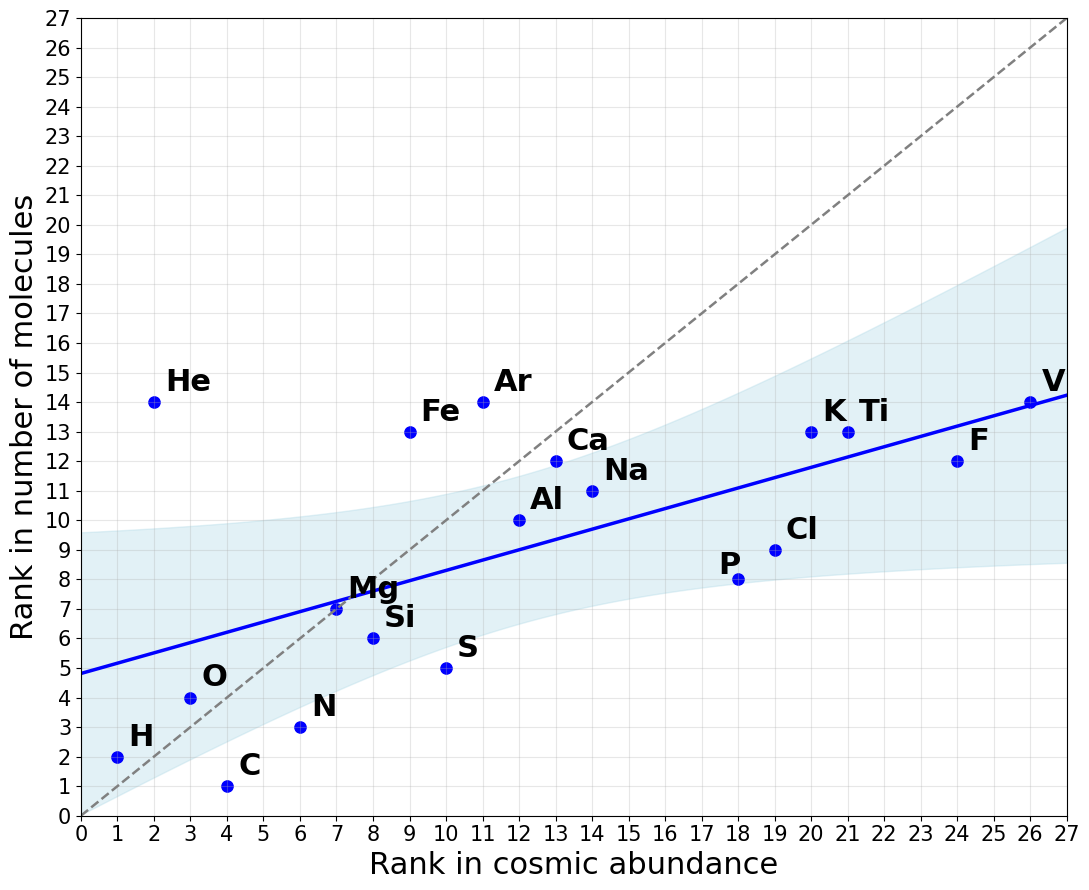}
    \vspace{-8mm}
    \caption{\small Comparison between the rank of chemical elements according to their cosmic abundance (\citealt{lodders2003}) and their rank in the number of interstellar molecules in which they are present (Fig. \ref{fig:histo-elements}). 
    The blue line shows the least-squares linear fit, and the shaded blue regions denotes the 3$\sigma$ uncertainty. The dashed black curve indicates the y=x relation.
    }
    \label{fig:ranks}
\end{wrapfigure}

These discoveries open a new path for the detection of species containing new chemical elements beyond the third row of the periodic table. 
%
%
%
%
Interestingly, there are still
 several cosmically abundant elements that 
have not been identified so far as constituents of interstellar molecules: Ne, Cu, Zn, Co, Mn, Cr, and Ni (Fig. \ref{fig:astromolecular-table}). Since there is an overall trend between the occurrence of elements in interstellar molecules and their cosmic abundance (Fig. \ref{fig:ranks}), molecules containing these elements are promising candidates for new detections. The elements beyond CHNOPS are also relevant agents in prebiotic chemistry, 
acting act as key catalysts and enabling synthesis of biomolecules
(\citealt{nuth2006,crabtree2026}).
As an example, 
Cu plays a pivotal catalytic role in driving the selective assembly of 
nucleotides, amino acids, and lipid precursors in laboratory prebiotic experiments (\citealt{patel2015}). Nowadays, very little is known about the 
reservoir of heavy element-bearing species in the ISM, and about how they can be incorporated to star-forming regions, planet forming-disks, protoplanets and planetesimals.

The recent revolution in the pace of new detected molecules suggests that we are just grasping the tip of the iceberg, both in terms of molecular size and chemical elemental diversity. 
Therefore, we are still far from a complete inventory.
In particular, {\bf molecules directly relevant to biochemistry such as nucleobases, lipids, amino acids} (Fig. \ref{fig:astro-to-bio}) have never been detected in the ISM, and only recently the first {\bf sugar} has been reported (\citealt{jimenez-serra2026}). Since these species are present in Solar System bodies (e.g., \citealt{glavin2025,furukawa2026}), 
is this because this prebiotic material is only synthesised within these bodies, e.g., through aqueous alteration (\citealt{takano2024})?
Or maybe the 
natal nebula of our Solar System was a unique and rare environment prone to the emergence of life? Or, alternatively, {\bf are the basic molecular precursors for life present elsewhere in the Galaxy?} 
In summary, the key 
open questions
are: 

\vspace{-1mm}

\begin{tcolorbox}[boxsep=3pt,left=0pt,right=0pt,top=1pt,bottom=1pt,colframe=arylideyellow,colback=arylideyellow]

\justify


{\bf

$\triangleright$  What are the frontiers of chemical complexity, both in terms of molecular size and elemental diversity, that can be achieved in the ISM? 

\hspace{-7mm}  $\triangleright$  How does chemical complexity emerge and evolve from the natal ISM to the subsequent phases of the formation of star and planetary systems?

\hspace{-7mm} $\triangleright$  What is the ISM chemistry's potential to provide building blocks to prebiotic chemistry?

}

\end{tcolorbox}

%
%
%


\vspace{-6mm}
\subsection{Current limitations and future challenges: Why ALMA 2040?}

\vspace{-2mm}

\begin{SCfigure*}
    \centering
\includegraphics[width=0.75\textwidth]{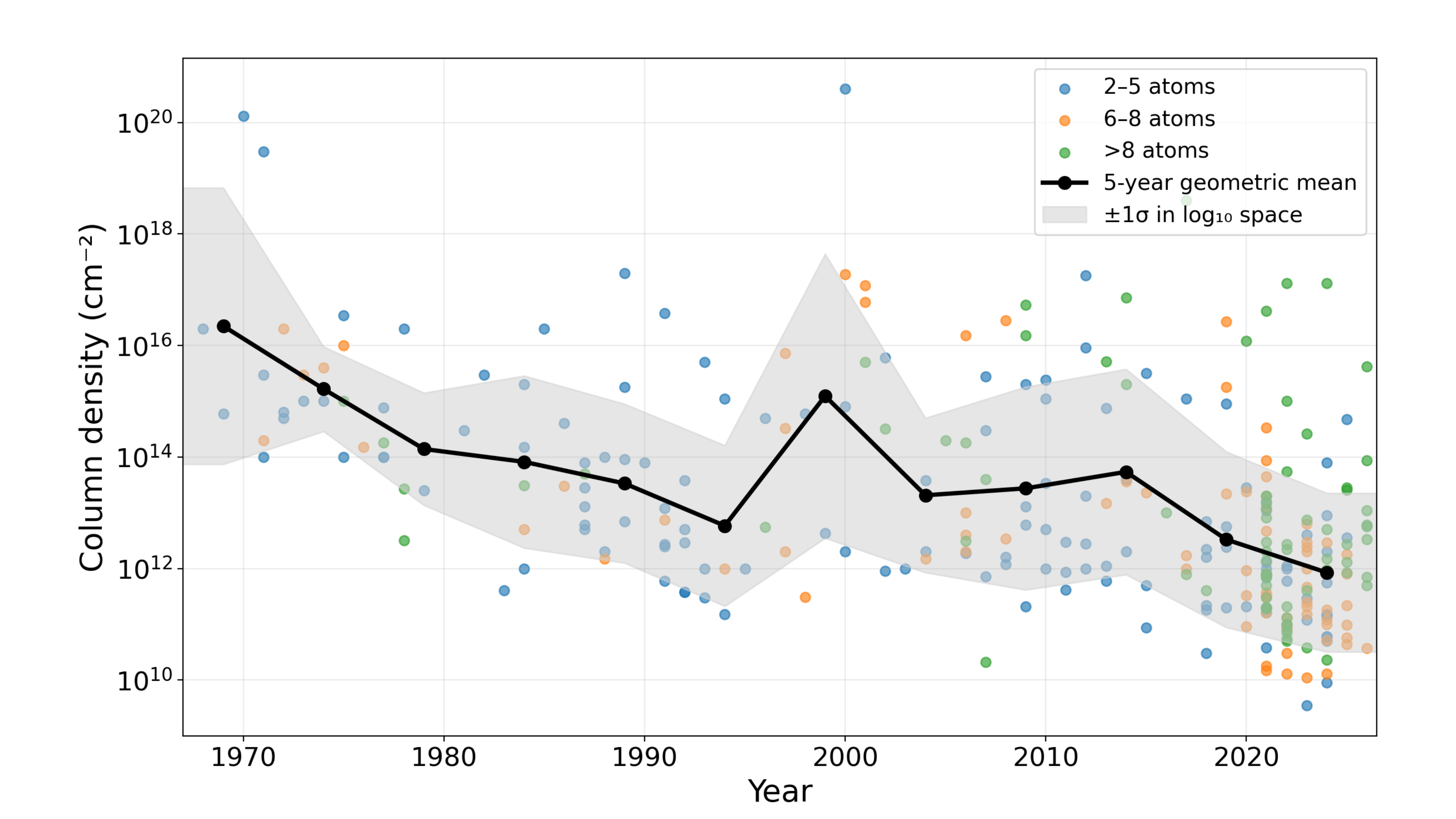}
    \hspace{-7mm}
    \vspace{-5mm}
    \caption{\small Evolution over the last half century of the values of the column density ($N$) of newly discovered interstellar molecules. The molecules are color-coded according to the number of constituent atoms. The solid black curve denotes the 5-year geometric mean in logarithmic scale, and the grey-shaded area represents the $\pm$1$\sigma$ uncertainty.
    }
    \label{fig:N_vs_yr}
\end{SCfigure*}

Only a new-generation (sub)millimeter interferometer (hereafter ALMA2040) 
will be transformative in this field, and will be able to answer all the previous questions. Such observatory 
must join
{\bf \color{Purple} ultra-high line sensitivity} at a wide variety of spatial scales, from compact ($\leq$1 $^{\prime\prime}$) to large structures ($\geq$20 $^{\prime\prime}$), and {\bf \color{Purple}  broadband instantaneous spectral coverage.}
%
%
To confirm the presence of a new molecule in the ISM, it is needed to detect several of its rotational transitions at significant signal-to-noise level, which
presents several observational challenges.
First, the expected line intensities 
are very weak, because the molecules have intrinsically low abundances, 
and in the case of complex molecules, also large partition functions that spread the population across many rotational levels.
Fig. \ref{fig:N_vs_yr} illustrates how our capacity of detecting new molecules with lower column densities ($N$) has improved in the last half century by around 4 orders of magnitude, especially for the larger species with $\geq$6 atoms. To allow the detection of new species with larger sizes - and thus prebiotic relevance - and containing heavier chemical elements, we will need to reach even lower values of molecular abundances of $\leq$10$^{-11}$, which in dense regions translates into typical column densities of $N<$ 10$^{11}-$10$^{12}$ cm$^{-2}$.

\begin{table}
\tabcolsep 1.5pt
\centering
\small
\caption{\small Requirements for ALMA2040 compared to those of the current single dishes, ALMA and ALMA-WSU.}
\begin{tabular}{ c c c c c c  }
\hline
 Subject & Requirement  & single dish & ALMA & ALMA-WSU &  ALMA2040\\
 \hline
 Angular resolution ($^{\prime\prime}$)  &  0.1 (cores) $-$ tens (clouds)   &  {\color{red}{\usym{2613}}} &{\color{ForestGreen}{$\checkmark$}}  & {\color{ForestGreen}{$\checkmark$}}& {\color{ForestGreen}{$\checkmark$}} \\
\hline
 Velocity resolution (km s$^{-1}$)  &  $\geq$ 0.05 & {\color{ForestGreen}{$\checkmark$}} & {\color{ForestGreen}{$\checkmark$}}  & {\color{ForestGreen}{$\checkmark$}} & {\color{ForestGreen}{$\checkmark$}} \\
\hline
  Frequency bands  &  1-10 & {\color{ForestGreen}{$\checkmark$}} &{\color{ForestGreen}{$\checkmark$}}  & {\color{ForestGreen}{$\checkmark$}} & {\color{ForestGreen}{$\checkmark$}} \\
\hline
{\bf  Instantaneous bandwidth (GHz) } & $>$512 (full ALMA bands) & {\color{red}{\usym{2613}}}  & {\color{red}{\usym{2613}}}   & {\color{red}{\usym{2613}}}  & {\color{ForestGreen}{$\checkmark$}}  \\
\hline
{\bf  Sensitivity (at all scales)  }& $\geq$10$\times$ ALMA & {\color{red}{\usym{2613}}}  & {\color{red}{\usym{2613}}}  & {\color{red}{\usym{2613}}} & {\color{ForestGreen}{$\checkmark$}} \\
 \hline

\hline
\end{tabular}
\label{tab:comparison}
\vspace{0mm}
\vspace*{-1.5ex}
\label{tab:alma2040-requirements}
\end{table}

The second challenge when observing chemically-rich interstellar sources 
is that they exhibit very crowded spectra, in which line contamination from more abundant (brighter) species might be a serious issue. These limitations can be addressed by performing ultra high-sensitivity broadband spectral surveys that provide very low noise levels and wide frequency coverage containing 
many transitions of the targeted molecules, which increases the probability of finding a good number of them free of line contamination.
%
%
Indeed, most of the new interstellar molecules discovered in the last years come from single-dish wideband and long-integration-time ($>$300-1500 hr) projects towards the cold molecular cloud TMC-1 by the QUIJOTE\footnote{Q-band Ultrasensitive Inspection Journey to the Obscure TMC-1
Environment.} (\citealt{cernicharo2021_indene}) and GOTHAM\footnote{GBT Observations of TMC-1: Hunting Aromatic Molecules.} (\citealt{mcguire2021}) teams, and towards the Galactic Center molecular cloud G+0.693-0.027 (\citealt{rivilla2019,rivilla2023,jimenez-serra2020,sanz-novo2023}).
%
%
However, these observations with large beams of tens of arcsec do not resolve the molecular emission down to smaller scales (Table \ref{tab:alma2040-requirements}). This spatial information is essential to understand the mechanisms governing the chemistry in each source, and also to trace the chemical thread across the different {\bf multi-scale evolutionary stages of star and planet formation}, from the cold natal molecular clouds (sizes of $\sim$1 pc), to protostellar cluster envelopes ($\sim$0.2 pc), to high-mass protostellar hot cores ($\sim$0.05 pc), to low-mass protostellar hot corinos and planet-forming disks ($\sim$50 au). 

%

%
%
%
%
%
%
%

\begin{wraptable}[14]{r}{0.55\textwidth}
\vspace{-3mm}
    \caption{\small Molecules whose first report in the ISM has been based on ALMA observations.}
\tabcolsep 0.15pt    
{\footnotesize    
    \centering
    \begin{tabular}{ l c }
    \hline
  Name (formula)  &  Reference  \\
    \hline                    
iso-propyl cyanide  (i-C$_3$H$_7$CN)  &  \citet{belloche2014}  \\
methyl chloride   (CH$_3$Cl)  & \citet{fayolle2017}   \\
methoxymethanol  (CH$_3$OCH$_2$OH)  & \citet{mcguire2017}   \\
nitrous acid   (HONO)  & \citet{coutens2019}  \\
glycolonitrile    (HOCH$_2$CN)  & \citet{zeng2019}    \\
hydroxyacetone      (CH$_3$COCH$_2$OH) & \citet{yan2020}   \\
isopropanol      (i-C$_3$H$_7$OH)  & \citet{Belloche2022}  \\
3-hydroxypropenal      (HOCHCHCHO)   & \citet{coutens2022}   \\
2-methoxyethanol       (CH$_3$OCH$_2$CH$_2$OH)  & \citet{Fried2024}  \\
protonated acetylene        (C$_2$H$_3^+$)  & \citet{muller2024}  \\
calcium sulfide          (CaS)  & \citet{Tasa-Chaveli2025}  \\
methyl carbamate        (CH$_3$OC(O)NH$_2$)  & \citet{duan2026}   \\
    \hline
    \end{tabular}
    \label{tab:alma-detections}
    }
\end{wraptable}

The advent of ALMA 15 yr ago offered the community the most sensitive (sub)millimeter interferometer to date, opening a privileged window to the {\it Molecular Universe}. While the advances that ALMA has promoted for 
astrochemistry 
are numerous and relevant, its impact in the quest of new molecules has been limited.
%
ALMA has provided the detection of 12 new species (Table \ref{tab:alma-detections}), which is a small fraction (7$\%$) of the total number discovered in these years ($>$160 from 2014). These molecules have been detected towards compact structures: hot cores in high-mass star-forming regions 
and hot corinos in low-mass star-forming regions, 
and 
towards an extragalactic molecular absorber toward a quasar (Table \ref{tab:alma-detections}).
Two main reasons explain why ALMA has not been a more powerful molecule detector (Table \ref{tab:alma2040-requirements}).
First, the {\bf limited instantaneous bandwidth} impose multiple frequency setups to cover broad frequency ranges, preventing to attain long integration times per setup. For example, the GUAPOS\footnote{G31.41+0.31 Unbiased ALMA sPectral Observational Survey} survey towards the G31.41+0.31 hot core needed 9 different setups to cover the full ALMA band 3 (\citealt{Mininni2020}); the PILS\footnote{Protostellar Interferometric Line Survey} survey used 18 setups to cover the range 329-362 GHz (\citealt{jorgensen2016}); and the  ALCHEMI\footnote{ALMA Comprehensive High-resolution Extragalactic Molecular Inventory} Large Program required 47 setups to cover the ALMA bands 3, 4, 5, 6 and 7 (\citealt{martin2021}). Namely, the completion of sensitive spectral surveys has been extremely time consuming, limiting the sensitivity achieved. On the other side, shorter projects based on a few setups are not able to detect of enough number of transitions, especially in hot 
sources, which suffer more line blending. For instance, the recent identification of CaS relies on a single unblended line (\citealt{Tasa-Chaveli2025}), and \citet{chunguo2026} only reported a tentative detection of glycolamide (NH$_2$COCH$_2$OH) in a hot core due to a low number of unblended transitions. 
%
%
This limitation will be partially mitigated with the ALMA Wideband Sensitivity Upgrade (WSU), which is planning to increase by a factor of 2$-$4 the instantaneous spectral coverage, up to 32 GHz (Table \ref{tab:alma2040-requirements}).

However, the {\bf true revolution for broadband ultra-sensitive spectral surveys will only arrive with a full instantaneous frequency coverage, across multiple ALMA bands ($>$512 GHz}, Table \ref{tab:alma2040-requirements}). This unprecedented capability will allow ultra-long integration times using a single spectral setup, providing access to multiple transitions of any kind of molecule (from small to large), and also a significant leap in sensitivity. It will not be needed to sacrifice spectral coverage for sensitivity, or viceversa. The instantaneous observation of all frequencies will exploit the advantages that each frequency range offers, mitigating its drawbacks.
The line intensities in hot core/corinos are brighter at higher frequencies (bands 5, 6 and 7)
, but the high line density in the spectra increases the blending problem; and additionally the emission can be even completely absorbed in the densest regions due to large dust opacity (\citealt{rivilla2017,desimone2020}). On the other side, the new ALMA low-frequency bands (1 and 2) show weaker emission, but they optimize the search for unblended transitions due to less crowded spectra (\citealt{beltran&rivilla2018}). A new facility covering, at the same time, all frequencies will allow us to avoid compromises that have limited so far our capacity to identify new molecules in a wide variety of environments.


%
The second main ALMA limitation is its
{\bf modest sensitivity to large scales ($\geq$10 $^{\prime\prime}$)}, which resulted into the filtering of large amounts of emission in sources such as clouds, clumps, accretion filaments or streamers. 
The limited 
brightness sensitivity of the Atacama Compact Array (ACA) and the total power (TP) antennas (which have only a $\sim$8$\%$ of the collecting area of the 12m-array) required very long integration times, around 10 times larger than that of the 12m-array,
to match image sensitivities at large scales (ALMA Technical Handbook). 
This, in practice, has made many projects unfeasible.
%
%
%
%
Thus, {\bf ALMA2040 design must 
integrate compact array and total power capabilities significantly improved, to
provide sensitive multi-scale high-fidelity imaging of all the sizes relevant in star and planet formation} (Table \ref{tab:comparison}).

\vspace{-5mm}
\section{The revolution of ALMA2040: Science cases 
}

\vspace{-2mm}
\subsection{Interstellar amino acids, nucleobases, sugars and lipids}

\vspace{-1mm}

\begin{figure*}[h!]
\centering
\includegraphics[width=1.0\textwidth]{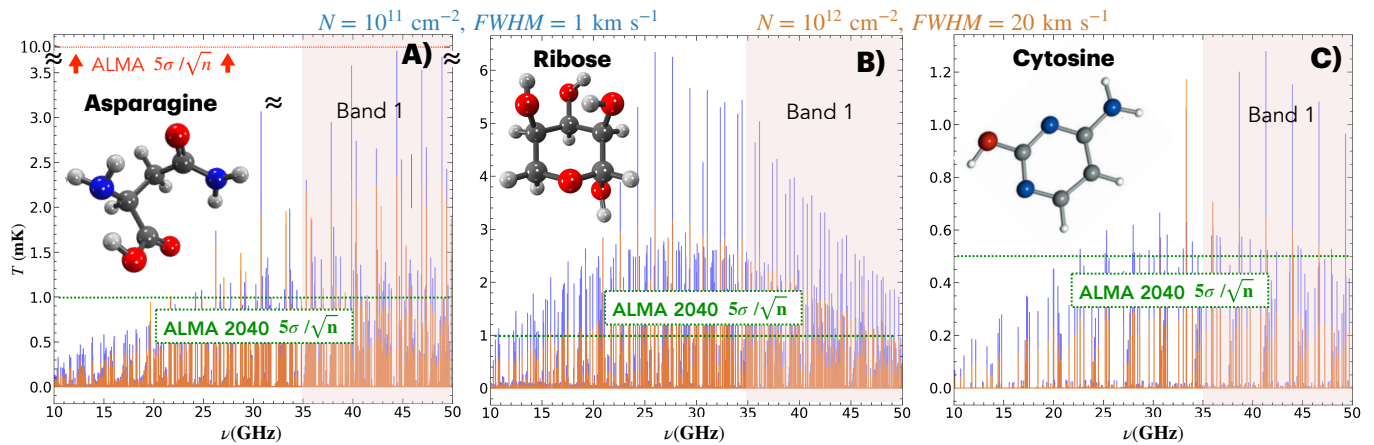}
\vspace{-8mm}
\caption{\small A) LTE simulated spectrum of the amino acid asparagine 
towards rotationally cold clouds with $T_{\rm ex}$\,=\,10\,K: a Galactic Center cloud (linewidth of 20\,km s$^{-1}$ and 
$N=10^{12}$ cm$^{-2}$; orange), and a Galactic disk cloud (1\,km s$^{-1}$ and $N=10^{11}$ cm$^{-2}$; blue).
The column densities correspond to abundances of $<$ 10$^{-11}$ for clouds with typical values of $N$(H$_2$) of $>10^{23}$ cm$^{-2}$ and $>10^{22}$ cm$^{-2}$, respectively. 
The velocity-integrated 5$\sigma$/$\sqrt{n}$ detection levels for ALMA2040 and current ALMA (the y-axis is broken to display purposes) for the same integration times are compared, where $n$ is the number of spectral channels of $\Delta v$ across the linewidth, and $\sigma$ is the $rms$ of the spectrum. ALMA2040 will be able to detect multiple transitions in integration times of $\sim$4 hr for the Galactic center cloud in a synthesized beam of $\sim$8$^{\prime\prime}$, and $\sim$25 hr for the Galactic disk cloud in a 
beam of $\sim$11$^{\prime\prime}$. 
B and C) Same as panel A, for ribose and cytosine, respectively. The ALMA velocity-integrated 5$\sigma$/$\sqrt{n}$ detection levels for ALMA are not shown because they fall out of the y-axis range presented. 
The lower expected line intensities of cytosine will require longer integration times of $\sim$20 hr in a 
beam of $\sim$8$^{\prime\prime}$ for the Galactic Center cloud, and $\sim$100 hr in a 
beam of $\sim11^{\prime\prime}$ for the Galactic disk cloud. 
} 
    \label{fig:prebiotic} 
\end{figure*}

ALMA2040 will be able to detect true building blocks of prebiotic chemsitry (Fig. \ref{fig:astro-to-bio}) in the ISM.
As examples, Fig. \ref{fig:prebiotic} shows the expected emission spectra of the amino acid asparagine (NH$_2$COCH$_2$CHNH$_2$COOH), the sugar ribose (C$_5$H$_{10}$O$_5$), and the nucleobase cytosine (C$_4$H$_5$N$_3$O), assuming Local Thermodynamic Equilibrium (LTE) conditions, using the available microwave rotational spectroscopy (\citealt{cabezas2012,Cocinero2012,Alonso2013}). 
%
%
We have considered typical physical conditions of clouds like TMC-1 and G+0.693-0.027, where the largest interstellar species have been detected. These sources have been particularly successful discovering new species because the excitation temperature  ($T_{\rm ex}$) of both clouds are low,  $\sim$10$\,$K (\citealt{agundez2023a}) for TMC-1 and 5$-$25 K for G+0.693-0.027 (\citealt{zeng2018,rivilla2022_nitriles}). Thus, only the lowest energy transitions are populated, providing spectra that are significantly less line-crowded compared to higher-$T_{\rm ex}$ sources (such as protostellar hot cores/corinos), which mitigates the line blending. 
For the LTE simulations we have used $T_{\rm ex}$=10 K, and FWHM of 1 km s$^{-1}$ and 20 km s$^{-1}$, respectively, and a low molecular abundance of 10$^{-11}$. 
The emission of these large species under cold excitation peaks in the ALMA band 1 (35$-$50 GHz). Fig. \ref{fig:prebiotic} shows that many transitions will be detectable at velocity-integrated significance level above 5 in 
beams of $\sim$8$^{\prime\prime}$-11$^{\prime\prime}$ using integration times of $\sim$4 to 100 hr (note that ALMA would require times longer by a factor of 100
, which are not feasible, and that single-dish facilities have not yet detected them despite ultra-long exposures of hundreds to thousands of hr). 

In the case of compact ($\sim$0.5$^{\prime\prime}$$-$1$^{\prime\prime}$) and hot ($T_{\rm ex}>$100 K) star-forming cores, the population peak shifts to higher frequencies, typically at bands 5, 6 and 7. Given that at those spectral ranges the impact of line blending is significant, it will be key to observe very broad bandwidths to increase the probability of identifying unblended transitions,
and to simultaneously cover lower frequencies, where the spectrum, although weaker, is less crowded. In addition, lower-frequency bands will also be suited to avoid the large dust opacities of hot cores/corinos at higher frequencies, which result in underestimation of the molecular abundances, or even completely quench the line emission (\citealt{rivilla2017,desimone2020}). Furthermore, the high angular resolution of ALMA2040, with beams $<$ 1$^{\prime\prime}$, will be able to spatially resolve the cores, which will provide narrower linewidths, and thus mitigation of the line blending.




\vspace{-4mm}
\subsection{Heavy-element-bearing molecules}
\vspace{-2mm}

\begin{figure*}[h!]
\centering
\includegraphics[width=0.975\textwidth]{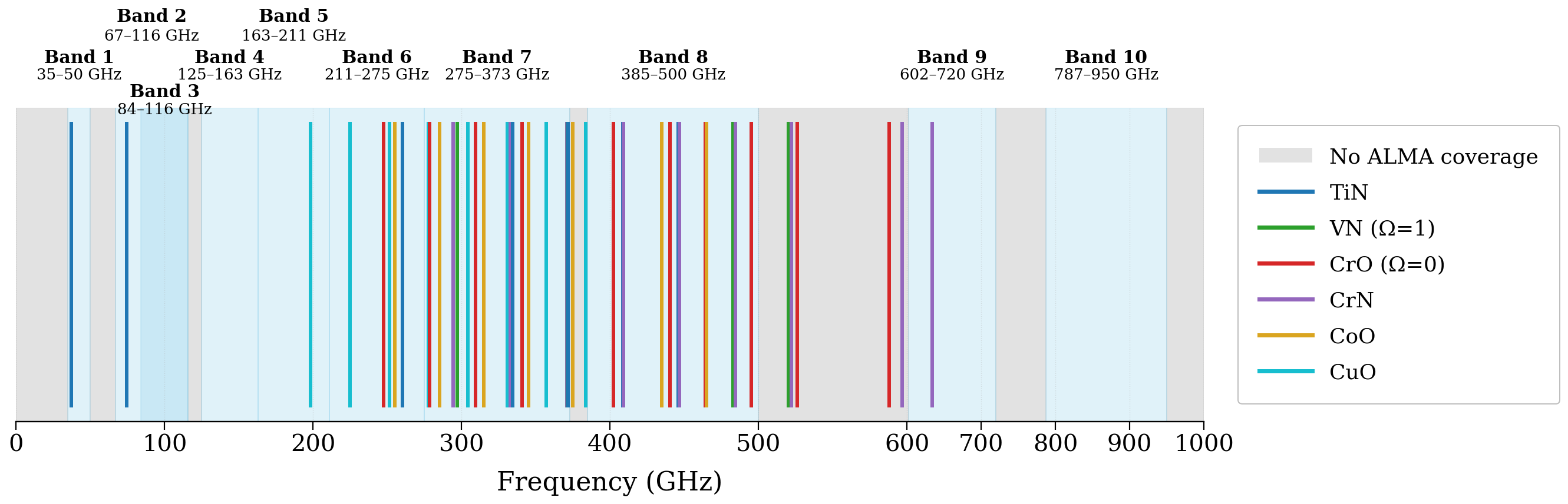}
\vspace{-4mm}
    \caption{\small Transitions of molecules not yet detected in the ISM, which contain heavy elements of the fourth row of the periodic table (Ti, V, Cr, Co and Cu; see Fig. \ref{fig:astromolecular-table}).} 
    \label{fig:metals} 
\end{figure*}

The increasing number of interstellar molecules with chemical elements beyond the second row of the periodic table (Figs. \ref{fig:astromolecular-table} and \ref{fig:histo-elements}), identified not only in circumstellar material, but also in diverse environment such as hot protostellar massive disks and in rotationally cold molecular clouds, is likely anticipating a new era of astrochemistry, in which heavier elements from refractory material of dust grains will be routinely present in new molecules. This new field of research will give rise to {\it Astromineralogy}. These species are unique tracers of the evaporation, destruction, and recondensation of refractory material. High angular resolution will be essential for distinguishing thermal evaporation in inner disks, sputtering in shocks, and chemical processing in disk winds. ALMA2040 will allow a quantitative investigation of the cycling of mineral-forming elements between dust and gas during star and planet formation.
We expect to detect new species containing elements that are already members of the {\it Astromolecular table of elements}, but also containing other cosmically abundant elements still missing (Fig. \ref{fig:astromolecular-table}). The emission of most of these molecules falls across the full ALMA2040 spectral coverage, and especially at band 6 and above, as shown in Fig. \ref{fig:metals}, which indicates the main emission lines of species not yet discovered in the ISM such as TiN, VN, CrO and CrN, CoO and CuO (spectroscopy from \citealt{steimle1997,namiki1998,namiki2001,sheridan2002,mcLamarrah2005,flory2008}). The robust confirmation of these species will require the detection of several transitions in different bands, and hence, the instantaneous ultra-broadband capability will be essential.

\vspace{-4mm}

\vspace{-1mm}
\section{ALMA2040: Technical Requirements}
\vspace{-0.3cm}



To address these key science questions ALMA2040 must fulfill the technical requirements summarized in Table \ref{tab:alma2040-requirements}, which
will significantly enhance current ALMA and near-future ALMA-WSU capabilities regarding multi-scale spectral line sensitivity
and instantaneous frequency coverage:

\vspace{-2mm}

\begin{tcolorbox}[boxsep=1pt,left=0pt,right=0pt,top=0.5pt,bottom=0.5pt,colframe=arylideyellow,colback=arylideyellow]
\justify
{ \hspace{5mm} \bf    $\geq$10$\times$ increase in spectral line sensitivity across a wide range of spatial scales} 
\end{tcolorbox}
\vspace{-2mm}
We require at least a $\geq$10$\times$ increase in line sensitivity compared to ALMA ($\geq$8$\times$ compared to anticipated ALMA-WSU capabilities), with improved image fidelity (through exquisite {\it uv}-coverage) across spatial scales from the smallest ($<$ 1$^{\prime\prime}$) to the most extended (tens of $^{\prime\prime}$), and excellent dynamic range.  
%
This will allow the detection of true prebiotic molecules such as those shown in Fig. \ref{fig:prebiotic}
and of species with heavy elements (Fig. \ref{fig:metals}).
This requirement can be achieved primarily by increasing the collecting area compared to ALMA/ALMA-WSU, including significantly improved short-baselines array and total power array with more antennas and/or larger dishes, and by disruptive developments in receiver technology.

\vspace{-1mm}
\begin{tcolorbox}[boxsep=1pt,left=0pt,right=0pt,top=1pt,bottom=1pt,colframe=arylideyellow,colback=arylideyellow]
\justify
{ \hspace{5mm} \bf  Ultra-broad instantaneous frequency coverage ($>$512 GHz)} 
\end{tcolorbox}
\vspace{-2mm}


ALMA2040 must be an efficient spectral survey machine that provides very wide instantaneous spectral coverage, even spanning the full frequency range between 35 and 950 GHz.





\vspace{-0.45cm}
\section{Synergies with other facilities}
\vspace{-0.3cm}


ALMA2040 will naturally benefit from synergies with other upcoming and planned facilities working at different wavelenghts:
{\bf  \color{purple} Current and upcoming facilities:}
{\color{purple} $\bullet$} {\it  \bf James Webb Space Telescope (JWST)}: infrared spectroscopic observations with NIRSpec and MIRI will allow 
the complete chemical inventory in, both, gas and ice to be assembled when combined with sub(millimeter) data.
{\color{purple} $\bullet$} {\it  \bf  Extremely Large Telescope (ELT)}: METIS will allow studies of interstellar ices with its IFU coverage of $\sim$3-5 $\mu$m at $<$1$^{\prime\prime}$ scales beyond the lifetime of the JWST.
{\color{purple} $\bullet$} {\it  \bf  Square Kilometer Array Observatory (SKAO)}: complementary low-frequency coverage in a large FoV.
%
%
%
{\bf \color{purple} Proposed middle-term planned facilities:}
{\color{purple} $\bullet$} {\it  \bf  Next generation Very Large Array (ngVLA)}: complementary observations of the northern hemisphere in a coherent frequency range (Band 1$\&$2); overlap of a fraction of the sky in which to perform complementary lower frequency band observations. 
{\color{purple} $\bullet$} {\it  \bf  Atacama Large Aperture Submillimeter Telescope (AtLAST)}: it can provide the zero spacing if  TP capabilities are not implemented in the first phase of ALMA2040, and wide-field and high-speed mapping to search for follow up targets with ALMA2040 at high spatial  resolution.

\vspace{-0.5cm}
\section*{Acknowledgements}
\vspace{-0.4cm}
This work was developed as part of the broader ALMA2040 community effort. We acknowledge the Lorentz Center and NOVA for supporting the community workshops and coordination activities that helped initiate and shape the ALMA2040 White Paper process.
We thank the two referees for their careful reading and their constructive suggestions.
V.M.R. acknowledges support from the grant PID2022-136814NB-I00 funded by the Spanish Ministry of Science, Innovation and Universities/State Agency of Research MICIU/AEI/10.13039/501100011033 and by ERDF, UE.

\vspace{2mm}

\bibliographystyle{apj_mod}
{\bibliography{bibliography,biblio}}

@ARTICLE{Sakai2008,
       author = {{Sakai}, Nami and {Sakai}, Takeshi and {Hirota}, Tomoya and {Yamamoto}, Satoshi},
        title = "{Abundant Carbon-Chain Molecules toward the Low-Mass Protostar IRAS 04368+2557 in L1527}",
      journal = {\apj},
         year = 2008,
        month = jan,
       volume = {672},
       number = {1},
        pages = {371-381},
          doi = {10.1086/523635},
       adsurl = {https://ui.adsabs.harvard.edu/abs/2008ApJ...672..371S}
}

@ARTICLE{Agundez2024,
       author = {{Ag{\'u}ndez}, M. and {Berm{\'u}dez}, C. and {Cabezas}, C. and {Molpeceres}, G. and {Endo}, Y. and {Marcelino}, N. and {Tercero}, B. and {Guillemin}, J. -C. and {de Vicente}, P. and {Cernicharo}, J.},
        title = "{The rich interstellar reservoir of dinitriles: Detection of malononitrile and maleonitrile in TMC-1}",
      journal = {\aap},
         year = 2024,
        month = aug,
       volume = {688},
          eid = {L31},
        pages = {L31},
          doi = {10.1051/0004-6361/202451525},
archivePrefix = {arXiv},
       eprint = {2408.02843},
 primaryClass = {astro-ph.GA},
       adsurl = {https://ui.adsabs.harvard.edu/abs/2024A&A...688L..31A}
}

@ARTICLE{Becker2019,
       author = {{Becker}, Sidney and {Feldmann}, Jonas and {Wiedemann}, Stefan and {Okamura}, Hidenori and {Schneider}, Christina and {Iwan}, Katharina and {Crisp}, Antony and {Rossa}, Martin and {Amatov}, Tynchtyk and {Carell}, Thomas},
        title = "{Unified prebiotically plausible synthesis of pyrimidine and purine RNA ribonucleotides}",
      journal = {Science},
         year = 2019,
        month = oct,
       volume = {366},
       number = {6461},
        pages = {76-82},
          doi = {10.1126/science.aax2747},
       adsurl = {https://ui.adsabs.harvard.edu/abs/2019Sci...366...76B}
}

@ARTICLE{Belloche2022,
       author = {{Belloche}, A. and {Garrod}, R.~T. and {Zingsheim}, O. and {M{\"u}ller}, H.~S.~P. and {Menten}, K.~M.},
        title = "{Interstellar detection and chemical modeling of iso-propanol and its normal isomer}",
      journal = {\aap},
         year = 2022,
        month = jun,
       volume = {662},
          eid = {A110},
        pages = {A110},
          doi = {10.1051/0004-6361/202243575},
archivePrefix = {arXiv},
       eprint = {2204.09912},
 primaryClass = {astro-ph.GA},
       adsurl = {https://ui.adsabs.harvard.edu/abs/2022A&A...662A.110B}
}

@ARTICLE{Benner2019,
       author = {{Benner}, Steven A. and {Kim}, Hyo-Joong and {Biondi}, Elisa},
        title = "{Prebiotic Chemistry that Could Not Not Have Happened}",
      journal = {Life},
         year = 2019,
        month = nov,
       volume = {9},
       number = {4},
        pages = {84},
          doi = {10.3390/life9040084},
       adsurl = {https://ui.adsabs.harvard.edu/abs/2019Life....9...84B}
}

@ARTICLE{cernicharo2024,
       author = {{Cernicharo}, J. and {Cabezas}, C. and {Fuentetaja}, R. and {Ag{\'u}ndez}, M. and {Tercero}, B. and {Janeiro}, J. and {Juanes}, M. and {Kaiser}, R.~I. and {Endo}, Y. and {Steber}, A.~L. and {P{\'e}rez}, D. and {P{\'e}rez}, C. and {Lesarri}, A. and {Marcelino}, N. and {de Vicente}, P.},
        title = "{Discovery of two cyano derivatives of acenaphthylene (C$_{12}$H$_{8}$) in TMC-1 with the QUIJOTE line survey}",
      journal = {\aap},
         year = 2024,
        month = oct,
       volume = {690},
          eid = {L13},
        pages = {L13},
          doi = {10.1051/0004-6361/202452196},
archivePrefix = {arXiv},
       eprint = {2409.19311},
 primaryClass = {astro-ph.GA},
       adsurl = {https://ui.adsabs.harvard.edu/abs/2024A&A...690L..13C}
}

@ARTICLE{Chyba&Sagan1992,
       author = {{Chyba}, Christopher and {Sagan}, Carl},
        title = "{Endogenous production, exogenous delivery and impact-shock synthesis of organic molecules: an inventory for the origins of life}",
      journal = {\nat},
         year = 1992,
        month = jan,
       volume = {355},
       number = {6356},
        pages = {125-132},
          doi = {10.1038/355125a0},
       adsurl = {https://ui.adsabs.harvard.edu/abs/1992Natur.355..125C}
}

@ARTICLE{Fried2024,
       author = {{Fried}, Zachary T.~P. and {El-Abd}, Samer J. and {Hays}, Brian M. and {Wenzel}, Gabi and {Byrne}, Alex N. and {Margul{\`e}s}, Laurent and {Motiyenko}, Roman A. and {Shipman}, Steven T. and {Horne}, Maria P. and {J{\o}rgensen}, Jes K. and {Brogan}, Crystal L. and {Hunter}, Todd R. and {Remijan}, Anthony J. and {Lipnicky}, Andrew and {Loomis}, Ryan A. and {McGuire}, Brett A.},
        title = "{Rotational Spectrum and First Interstellar Detection of 2-methoxyethanol Using ALMA Observations of NGC 6334I}",
      journal = {\apjl},
         year = 2024,
        month = apr,
       volume = {965},
       number = {2},
          eid = {L23},
        pages = {L23},
          doi = {10.3847/2041-8213/ad37ff},
archivePrefix = {arXiv},
       eprint = {2403.17341},
 primaryClass = {astro-ph.GA},
       adsurl = {https://ui.adsabs.harvard.edu/abs/2024ApJ...965L..23F}
}

@ARTICLE{McGuire2020,
       author = {{McGuire}, Brett A. and {Burkhardt}, Andrew M. and {Loomis}, Ryan A. and {Shingledecker}, Christopher N. and {Kelvin Lee}, Kin Long and {Charnley}, Steven B. and {Cordiner}, Martin A. and {Herbst}, Eric and {Kalenskii}, Sergei and {Momjian}, Emmanuel and {Willis}, Eric R. and {Xue}, Ci and {Remijan}, Anthony J. and {McCarthy}, Michael C.},
        title = "{Early Science from GOTHAM: Project Overview, Methods, and the Detection of Interstellar Propargyl Cyanide (HCCCH$_{2}$CN) in TMC-1}",
      journal = {\apjl},
         year = 2020,
        month = sep,
       volume = {900},
       number = {1},
          eid = {L10},
        pages = {L10},
          doi = {10.3847/2041-8213/aba632},
archivePrefix = {arXiv},
       eprint = {2008.12349},
 primaryClass = {astro-ph.GA},
       adsurl = {https://ui.adsabs.harvard.edu/abs/2020ApJ...900L..10M}
}

@ARTICLE{McGuire2022,
       author = {{McGuire}, Brett A.},
        title = "{2021 Census of Interstellar, Circumstellar, Extragalactic, Protoplanetary Disk, and Exoplanetary Molecules}",
      journal = {\apjs},
         year = 2022,
        month = apr,
       volume = {259},
       number = {2},
          eid = {30},
        pages = {30},
          doi = {10.3847/1538-4365/ac2a48},
archivePrefix = {arXiv},
       eprint = {2109.13848},
 primaryClass = {astro-ph.GA},
       adsurl = {https://ui.adsabs.harvard.edu/abs/2022ApJS..259...30M}
}

@ARTICLE{Patel2015,
       author = {{Patel}, Bhavesh H. and {Percivalle}, Claudia and {Ritson}, Dougal J. and {Duffy}, Colm D. and {Sutherland}, John D.},
        title = "{Common origins of RNA, protein and lipid precursors in a cyanosulfidic protometabolism}",
      journal = {Nature Chemistry},
         year = 2015,
        month = apr,
       volume = {7},
       number = {4},
        pages = {301-307},
          doi = {10.1038/nchem.2202},
       adsurl = {https://ui.adsabs.harvard.edu/abs/2015NatCh...7..301P}
}

@ARTICLE{Powner2009,
       author = {{Powner}, Matthew W. and {Gerland}, B{\'e}atrice and {Sutherland}, John D.},
        title = "{Synthesis of activated pyrimidine ribonucleotides in prebiotically plausible conditions}",
      journal = {\nat},
         year = 2009,
        month = may,
       volume = {459},
       number = {7244},
        pages = {239-242},
          doi = {10.1038/nature08013},
       adsurl = {https://ui.adsabs.harvard.edu/abs/2009Natur.459..239P}
}

@ARTICLE{Rey-Montejo2024,
       author = {{Rey-Montejo}, Marta and {Jim{\'e}nez-Serra}, Izaskun and {Mart{\'\i}n-Pintado}, Jes{\'u}s and {Rivilla}, V{\'\i}ctor M. and {Meg{\'\i}as}, Andr{\'e}s and {San Andr{\'e}s}, David and {Sanz-Novo}, Miguel and {Colzi}, Laura and {Zeng}, Shaoshan and {L{\'o}pez-Gallifa}, {\'A}lvaro and {Mart{\'\i}nez-Henares}, Antonio and {Mart{\'\i}n}, Sergio and {Tercero}, Bel{\'e}n and {de Vicente}, Pablo and {Requena-Torres}, Miguel},
        title = "{Discovery of MgS and NaS in the Interstellar Medium and Tentative Detection of CaO}",
      journal = {\apj},
         year = 2024,
        month = nov,
       volume = {975},
       number = {2},
          eid = {174},
        pages = {174},
          doi = {10.3847/1538-4357/ad736e},
archivePrefix = {arXiv},
       eprint = {2407.07693},
 primaryClass = {astro-ph.GA},
       adsurl = {https://ui.adsabs.harvard.edu/abs/2024ApJ...975..174R}
}

@ARTICLE{Rivilla2020b,
       author = {{Rivilla}, V{\'\i}ctor M. and {Mart{\'\i}n-Pintado}, Jes{\'u}s and {Jim{\'e}nez-Serra}, Izaskun and {Mart{\'i}n}, Sergio and {Rodr{\'i}guez-Almeida}, Lucas F. and {Requena-Torres}, Miguel A. and {Rico-Villas}, Fernando and {Zeng}, Shaoshan and {Briones}, Carlos},
        title = "{Prebiotic Precursors of the Primordial RNA World in Space: Detection of NH$_{2}$OH}",
      journal = {\apjl},
         year = 2020,
        month = aug,
       volume = {899},
       number = {2},
          eid = {L28},
        pages = {L28},
          doi = {10.3847/2041-8213/abac55},
archivePrefix = {arXiv},
       eprint = {2008.00228},
 primaryClass = {astro-ph.GA},
       adsurl = {https://ui.adsabs.harvard.edu/abs/2020ApJ...899L..28R}
}

@book{crabtree2026,
  title={The organometallic chemistry of the transition metals},
  author={{Crabtree}, Robert H},
  year={2026},
  publisher={John Wiley \& Sons}
}

@article{nuth2006,
  title={Chemical processes in the interstellar medium: Source of the gas and dust in the primitive solar nebula},
  author={{Nuth}, Joseph A and Charnley, Steven B and Johnson, Natasha M},
  journal={Meteorites and the early solar system II},
  volume={2},
  pages={147--167},
  year={2006},
  publisher={this volume. Univ. of Arizona, Tucson}
}

@ARTICLE{rivilla2022_nitriles,
       author = {{Rivilla}, V{\'\i}ctor M. and {Jim{\'e}nez-Serra}, Izaskun and {Mart{\'\i}n-Pintado}, Jes{\'u}s and {Colzi}, Laura and {Tercero}, Bel{\'e}n and {de Vicente}, Pablo and {Zeng}, Shaoshan and {Mart{\'\i}n}, Sergio and {Garc{\'\i}a de la Concepci{\'o}n}, Juan and {Bizzocchi}, Luca and {Melosso}, Mattia and {Rico-Villas}, Fernando and {Requena-Torres}, Miguel A.},
        title = "{Molecular Precursors of the RNA-World in Space: New Nitriles in the G+0.693{\ensuremath{-}}0.027 Molecular Cloud}",
      journal = {Frontiers in Astronomy and Space Sciences},
         year = 2022,
        month = jul,
       volume = {9},
          eid = {876870},
        pages = {876870},
          doi = {10.3389/fspas.2022.876870},
archivePrefix = {arXiv},
       eprint = {2206.01053},
 primaryClass = {astro-ph.GA},
       adsurl = {https://ui.adsabs.harvard.edu/abs/2022FrASS...9.6870R}
}

@ARTICLE{Sanz-Novo2024b,
       author = {{Sanz-Novo}, Miguel and {Rivilla}, V{\'\i}ctor M. and {M{\"u}ller}, Holger S.~P. and {Jim{\'e}nez-Serra}, Izaskun and {Mart{\'\i}n-Pintado}, Jes{\'u}s and {Colzi}, Laura and {Zeng}, Shaoshan and {Meg{\'\i}as}, Andr{\'e}s and {L{\'o}pez-Gallifa}, {\'A}lvaro and {Mart{\'\i}nez-Henares}, Antonio and {Tercero}, Bel{\'e}n and {de Vicente}, Pablo and {San Andr{\'e}s}, David and {Mart{\'\i}n}, Sergio and {Requena-Torres}, Miguel A.},
        title = "{Discovery of Thionylimide, HNSO, in Space: The first N-, S-, and O-bearing Interstellar Molecule}",
      journal = {ApJLetters},
         year = 2024,
        month = apr,
       volume = {965},
       number = {2},
          eid = {L26},
        pages = {L26},
          doi = {10.3847/2041-8213/ad3945},
archivePrefix = {arXiv},
       eprint = {2404.01044},
 primaryClass = {astro-ph.GA},
       adsurl = {https://ui.adsabs.harvard.edu/abs/2024ApJ...965L..26S}
}

@ARTICLE{agundez2023a,
       author = {{Ag{\'u}ndez}, M. and {Marcelino}, N. and {Tercero}, B. and {Cernicharo}, J.},
        title = "{Aromatic cycles are widespread in cold clouds}",
      journal = {\aap},
         year = 2023,
        month = sep,
       volume = {677},
          eid = {L13},
        pages = {L13},
          doi = {10.1051/0004-6361/202347524},
archivePrefix = {arXiv},
       eprint = {2308.15951},
 primaryClass = {astro-ph.GA},
       adsurl = {https://ui.adsabs.harvard.edu/abs/2023A&A...677L..13A}
}

@ARTICLE{cernicharo2021_indene,
       author = {{Cernicharo}, J. and {Ag{\'u}ndez}, M. and {Cabezas}, C. and {Tercero}, B. and {Marcelino}, N. and {Pardo}, J.~R. and {de Vicente}, P.},
        title = "{Pure hydrocarbon cycles in TMC-1: Discovery of ethynyl cyclopropenylidene, cyclopentadiene, and indene}",
      journal = {\aap},
         year = 2021,
        month = may,
       volume = {649},
          eid = {L15},
        pages = {L15},
          doi = {10.1051/0004-6361/202141156},
archivePrefix = {arXiv},
       eprint = {2104.13991},
 primaryClass = {astro-ph.GA},
       adsurl = {https://ui.adsabs.harvard.edu/abs/2021A&A...649L..15C}
}

@ARTICLE{mcguire2021,
       author = {{McGuire}, Brett A. and {Loomis}, Ryan A. and {Burkhardt}, Andrew M. and {Lee}, Kin Long Kelvin and {Shingledecker}, Christopher N. and {Charnley}, Steven B. and {Cooke}, Ilsa R. and {Cordiner}, Martin A. and {Herbst}, Eric and {Kalenskii}, Sergei and {Siebert}, Mark A. and {Willis}, Eric R. and {Xue}, Ci and {Remijan}, Anthony J. and {McCarthy}, Michael C.},
        title = "{Detection of two interstellar polycyclic aromatic hydrocarbons via spectral matched filtering}",
      journal = {Science},
         year = 2021,
        month = mar,
       volume = {371},
       number = {6535},
        pages = {1265-1269},
          doi = {10.1126/science.abb7535},
archivePrefix = {arXiv},
       eprint = {2103.09984},
 primaryClass = {astro-ph.GA},
       adsurl = {https://ui.adsabs.harvard.edu/abs/2021Sci...371.1265M}
}

@ARTICLE{wenzel2025,
       author = {{Wenzel}, Gabi and {Gong}, Siyuan and {Xue}, Ci and {Changala}, P. Bryan and {Holdren}, Martin S. and {Speak}, Thomas H. and {Stewart}, D. Archie and {Fried}, Zachary T.~P. and {Willis}, Reace H.~J. and {Bergin}, Edwin A. and {Burkhardt}, Andrew M. and {Byrne}, Alex N. and {Charnley}, Steven B. and {Lipnicky}, Andrew and {Loomis}, Ryan A. and {Shingledecker}, Christopher N. and {Cooke}, Ilsa R. and {McCarthy}, Michael C. and {Remijan}, Anthony J. and {Wendlandt}, Alison E. and {McGuire}, Brett A.},
        title = "{Discovery of the Seven-ring Polycyclic Aromatic Hydrocarbon Cyanocoronene (C$_{24}$H$_{11}$CN) in GOTHAM Observations of TMC-1}",
      journal = {\apjl},
         year = 2025,
        month = may,
       volume = {984},
       number = {1},
          eid = {L36},
        pages = {L36},
          doi = {10.3847/2041-8213/adc911},
archivePrefix = {arXiv},
       eprint = {2504.05232},
 primaryClass = {astro-ph.GA},
       adsurl = {https://ui.adsabs.harvard.edu/abs/2025ApJ...984L..36W}
}

@INBOOK{pizzarello2006,
       author = {{Pizzarello}, S. and {Cooper}, G.~W. and {Flynn}, G.~J.},
        title = "{The Nature and Distribution of the Organic Material in Carbonaceous Chondrites and Interplanetary Dust Particles}",
    booktitle = {Meteorites and the Early Solar System II},
         year = 2006,
       editor = {{Lauretta}, Dante S. and {McSween}, Harry Y.},
        pages = {625},
       adsurl = {https://ui.adsabs.harvard.edu/abs/2006mess.book..625P}
}

@article{cooper2001,
  title={Carbonaceous meteorites as a source of sugar-related organic compounds for the early Earth},
  author={{Cooper}, George and Kimmich, Novelle and Belisle, Warren and Sarinana, Josh and Brabham, Katrina and Garrel, Laurence},
  journal={Nature},
  volume={414},
  number={6866},
  pages={879--883},
  year={2001},
  publisher={Nature Publishing Group}
}

@ARTICLE{desimone2020,
       author = {{De Simone}, Marta and {Ceccarelli}, Cecilia and {Codella}, Claudio and {Svoboda}, Brian E. and {Chandler}, Claire and {Bouvier}, Mathilde and {Yamamoto}, Satoshi and {Sakai}, Nami and {Caselli}, Paola and {Favre}, Cecile and {Loinard}, Laurent and {Lefloch}, Bertrand and {Liu}, Hauyu Baobab and {L{\'o}pez-Sepulcre}, Ana and {Pineda}, Jaime E. and {Taquet}, Vianney and {Testi}, Leonardo},
        title = "{Hot Corinos Chemical Diversity: Myth or Reality?}",
      journal = {\apjl},
         year = 2020,
        month = jun,
       volume = {896},
       number = {1},
          eid = {L3},
        pages = {L3},
          doi = {10.3847/2041-8213/ab8d41},
archivePrefix = {arXiv},
       eprint = {2006.04484},
 primaryClass = {astro-ph.SR},
       adsurl = {https://ui.adsabs.harvard.edu/abs/2020ApJ...896L...3D}
}

@ARTICLE{agundez2026,
       author = {{Ag{\'u}ndez}, Marcelino and {Cernicharo}, Jos{\'e}},
        title = "{TMC-1: Probing the Onset of Chemical Complexity in Space}",
      journal = {ACS Earth and Space Chemistry},
         year = 2026,
        month = may,
       volume = {10},
       number = {5},
        pages = {1345-1359},
          doi = {10.1021/acsearthspacechem.6c00026},
archivePrefix = {arXiv},
       eprint = {2604.26439},
 primaryClass = {astro-ph.GA},
       adsurl = {https://ui.adsabs.harvard.edu/abs/2026ESC....10.1345A}
}

@ARTICLE{glavin2025,
       author = {{Glavin}, Daniel P. and {Dworkin}, Jason P. and {Alexander}, Conel M. O'D. and {Aponte}, Jos{\'e} C. and {Baczynski}, Allison A. and {Barnes}, Jessica J. and {Bechtel}, Hans A. and {Berger}, Eve L. and {Burton}, Aaron S. and {Caselli}, Paola and {Chung}, Angela H. and {Clemett}, Simon J. and {Cody}, George D. and {Dominguez}, Gerardo and {Elsila}, Jamie E. and {Farnsworth}, Kendra K. and {Foustoukos}, Dionysis I. and {Freeman}, Katherine H. and {Furukawa}, Yoshihiro and {Gainsforth}, Zack and {Graham}, Heather V. and {Grassi}, Tommaso and {Giuliano}, Barbara Michela and {Hamilton}, Victoria E. and {Haenecour}, Pierre and {Heck}, Philipp R. and {Hofmann}, Amy E. and {House}, Christopher H. and {Huang}, Yongsong and {Kaplan}, Hannah H. and {Keller}, Lindsay P. and {Kim}, Bumsoo and {Koga}, Toshiki and {Liss}, Michael and {McLain}, Hannah L. and {Marcus}, Matthew A. and {Matney}, Mila and {McCoy}, Timothy J. and {McIntosh}, Oph{\'e}lie M. and {Mojarro}, Angel and {Naraoka}, Hiroshi and {Nguyen}, Ann N. and {Nuevo}, Michel and {Nuth}, Joseph A. and {Oba}, Yasuhiro and {Parker}, Eric T. and {Peretyazhko}, Tanya S. and {Sandford}, Scott A. and {Santos}, Ewerton and {Schmitt-Kopplin}, Philippe and {Seguin}, Frederic and {Simkus}, Danielle N. and {Shahid}, Anique and {Takano}, Yoshinori and {Thomas-Keprta}, Kathie L. and {Tripathi}, Havishk and {Weiss}, Gabriella and {Zheng}, Yuke and {Lunning}, Nicole G. and {Righter}, Kevin and {Connolly}, Harold C. and {Lauretta}, Dante S.},
        title = "{Abundant ammonia and nitrogen-rich soluble organic matter in samples from asteroid (101955) Bennu}",
      journal = {Nature Astronomy},
         year = 2025,
        month = feb,
       volume = {9},
        pages = {199-210},
          doi = {10.1038/s41550-024-02472-9},
       adsurl = {https://ui.adsabs.harvard.edu/abs/2025NatAs...9..199G}
}

@ARTICLE{furukawa2026,
       author = {{Furukawa}, Yoshihiro and {Sunami}, Sako and {Takano}, Yoshinori and {Koga}, Toshiki and {Hirakawa}, Yuta and {Oba}, Yasuhiro and {Naraoka}, Hiroshi and {Saigusa}, Daisuke and {Yoshikawa}, Takaaki and {Tanaka}, Satoru and {Glavin}, Daniel P. and {Dworkin}, Jason P. and {Connolly}, Jr., Harold C. and {Lauretta}, Dante S.},
        title = "{Bio-essential sugars in samples from asteroid Bennu}",
      journal = {Nature Geoscience},
         year = 2026,
        month = jan,
       volume = {19},
       number = {1},
        pages = {19-24},
          doi = {10.1038/s41561-025-01838-6},
       adsurl = {https://ui.adsabs.harvard.edu/abs/2026NatGe..19...19F}
}

@ARTICLE{lopez-gallifa2025,
       author = {{L{\'o}pez-Gallifa}, {\'A}. and {Rivilla}, V.~M. and {Beltr{\'a}n}, M.~T. and {Colzi}, L. and {Fontani}, F. and {S{\'a}nchez-Monge}, {\'A}. and {Mininni}, C. and {Cesaroni}, R. and {Jim{\'e}nez-Serra}, I. and {Viti}, S. and {Lorenzani}, A.},
        title = "{The GUAPOS project: VI. The chemical inventory of shocked gas}",
      journal = {\aap},
         year = 2025,
        month = dec,
       volume = {704},
          eid = {A288},
        pages = {A288},
          doi = {10.1051/0004-6361/202556837},
archivePrefix = {arXiv},
       eprint = {2509.16094},
 primaryClass = {astro-ph.GA},
       adsurl = {https://ui.adsabs.harvard.edu/abs/2025A&A...704A.288L}
}

@ARTICLE{Oba2023,
       author = {{Oba}, Yasuhiro and {Koga}, Toshiki and {Takano}, Yoshinori and {Ogawa}, Nanako O. and {Ohkouchi}, Naohiko and {Sasaki}, Kazunori and {Sato}, Hajime and {Glavin}, Daniel P. and {Dworkin}, Jason P. and {Naraoka}, Hiroshi and {Tachibana}, Shogo and {Yurimoto}, Hisayoshi and {Nakamura}, Tomoki and {Noguchi}, Takaaki and {Okazaki}, Ryuji and {Yabuta}, Hikaru and {Sakamoto}, Kanako and {Yada}, Toru and {Nishimura}, Masahiro and {Nakato}, Aiko and {Miyazaki}, Akiko and {Yogata}, Kasumi and {Abe}, Masanao and {Okada}, Tatsuaki and {Usui}, Tomohiro and {Yoshikawa}, Makoto and {Saiki}, Takanao and {Tanaka}, Satoshi and {Terui}, Fuyuto and {Nakazawa}, Satoru and {Watanabe}, Sei-ichiro and {Tsuda}, Yuichi and {Hayabusa2-initial-analysis SOM team}},
        title = "{Uracil in the carbonaceous asteroid (162173) Ryugu}",
      journal = {Nature Communications},
         year = 2023,
        month = mar,
       volume = {14},
          eid = {1292},
        pages = {1292},
          doi = {10.1038/s41467-023-36904-3},
       adsurl = {https://ui.adsabs.harvard.edu/abs/2023NatCo..14.1292O}
}

@INPROCEEDINGS{Dutrey2014,
       author = {{Dutrey}, A. and {Semenov}, D. and {Chapillon}, E. and {Gorti}, U. and {Guilloteau}, S. and {Hersant}, F. and {Hogerheijde}, M. and {Hughes}, M. and {Meeus}, G. and {Nomura}, H. and {Pi{\'e}tu}, V. and {Qi}, C. and {Wakelam}, V.},
        title = "{Physical and Chemical Structure of Planet-Forming Disks Probed by Millimeter Observations and Modeling}",
    booktitle = {Protostars and Planets VI},
         year = 2014,
       editor = {{Beuther}, Henrik and {Klessen}, Ralf S. and {Dullemond}, Cornelis P. and {Henning}, Thomas},
        month = jan,
        pages = {317-338},
          doi = {10.2458/azu_uapress_9780816531240-ch014},
archivePrefix = {arXiv},
       eprint = {1402.3503},
 primaryClass = {astro-ph.GA},
       adsurl = {https://ui.adsabs.harvard.edu/abs/2014prpl.conf..317D}
}

@ARTICLE{Booth2021,
       author = {{Booth}, Alice S. and {Walsh}, Catherine and {Terwisscha van Scheltinga}, Jeroen and {van Dishoeck}, Ewine F. and {Ilee}, John D. and {Hogerheijde}, Michiel R. and {Kama}, Mihkel and {Nomura}, Hideko},
        title = "{An inherited complex organic molecule reservoir in a warm planet-hosting disk}",
      journal = {Nature Astronomy},
         year = 2021,
        month = jan,
       volume = {5},
        pages = {684-690},
          doi = {10.1038/s41550-021-01352-w},
archivePrefix = {arXiv},
       eprint = {2104.08348},
 primaryClass = {astro-ph.EP},
       adsurl = {https://ui.adsabs.harvard.edu/abs/2021NatAs...5..684B}
}

@ARTICLE{Brunken2022,
       author = {{Brunken}, Nashanty G.~C. and {Booth}, Alice S. and {Leemker}, Margot and {Nazari}, Pooneh and {van der Marel}, Nienke and {van Dishoeck}, Ewine F.},
        title = "{A major asymmetric ice trap in a planet-forming disk. III. First detection of dimethyl ether}",
      journal = {\aap},
         year = 2022,
        month = mar,
       volume = {659},
          eid = {A29},
        pages = {A29},
          doi = {10.1051/0004-6361/202142981},
archivePrefix = {arXiv},
       eprint = {2203.02936},
 primaryClass = {astro-ph.EP},
       adsurl = {https://ui.adsabs.harvard.edu/abs/2022A&A...659A..29B}
}

@ARTICLE{Lefloch2017,
       author = {{Lefloch}, Bertrand and {Ceccarelli}, C. and {Codella}, C. and {Favre}, C. and {Podio}, L. and {Vastel}, C. and {Viti}, S. and {Bachiller}, R.},
        title = "{L1157-B1, a factory of complex organic molecules in a solar-type star-forming region}",
      journal = {\mnras},
         year = 2017,
        month = jul,
       volume = {469},
       number = {1},
        pages = {L73-L77},
          doi = {10.1093/mnrasl/slx050},
archivePrefix = {arXiv},
       eprint = {1704.04646},
 primaryClass = {astro-ph.GA},
       adsurl = {https://ui.adsabs.harvard.edu/abs/2017MNRAS.469L..73L}
}

@ARTICLE{bachiller1997,
       author = {{Bachiller}, R. and {P{\'e}rez Guti{\'e}rrez}, M.},
        title = "{Shock Chemistry in the Young Bipolar Outflow L1157}",
      journal = {\apjl},
         year = 1997,
        month = sep,
       volume = {487},
       number = {1},
        pages = {L93-L96},
          doi = {10.1086/310877},
       adsurl = {https://ui.adsabs.harvard.edu/abs/1997ApJ...487L..93B}
}

@ARTICLE{pratap1997,
       author = {{Pratap}, P. and {Dickens}, J.~E. and {Snell}, R.~L. and {Miralles}, M.~P. and {Bergin}, E.~A. and {Irvine}, W.~M. and {Schloerb}, F.~P.},
        title = "{A Study of the Physics and Chemistry of TMC-1}",
      journal = {\apj},
         year = 1997,
        month = sep,
       volume = {486},
       number = {2},
        pages = {862-885},
          doi = {10.1086/304553},
       adsurl = {https://ui.adsabs.harvard.edu/abs/1997ApJ...486..862P}
}

@ARTICLE{jansen1995,
       author = {{Jansen}, D.~J. and {Spaans}, M. and {Hogerheijde}, M.~R. and {van Dishoeck}, E.~F.},
        title = "{Millimeter and submillimeter observations of the Orion Bar. II. Chemical models.}",
      journal = {\aap},
         year = 1995,
        month = nov,
       volume = {303},
        pages = {541},
       adsurl = {https://ui.adsabs.harvard.edu/abs/1995A&A...303..541J}
}

@ARTICLE{lucas1996,
       author = {{Lucas}, R. and {Liszt}, H.},
        title = "{The Plateau de Bure survey of galactic {\ensuremath{\lambda}}3mm HCO$^{+}$ absorption toward compact extragalactic continuum sources.}",
      journal = {\aap},
         year = 1996,
        month = mar,
       volume = {307},
        pages = {237},
       adsurl = {https://ui.adsabs.harvard.edu/abs/1996A&A...307..237L}
}

@ARTICLE{ziurys2006,
       author = {{Ziurys}, Lucy M.},
        title = "{Interstellar Chemistry Special Feature: The chemistry in circumstellar envelopes of evolved stars: Following the origin of the elements to the origin of life}",
      journal = {Proceedings of the National Academy of Science},
         year = 2006,
        month = aug,
       volume = {103},
       number = {33},
        pages = {12274-12279},
          doi = {10.1073/pnas.0602277103},
       adsurl = {https://ui.adsabs.harvard.edu/abs/2006PNAS..10312274Z}
}

@ARTICLE{cernicharo2000,
       author = {{Cernicharo}, J. and {Gu{\'e}lin}, M. and {Kahane}, C.},
        title = "{A lambda 2 mm molecular line survey of the C-star envelope IRC+10216}",
      journal = {\aaps},
         year = 2000,
        month = mar,
       volume = {142},
        pages = {181-215},
          doi = {10.1051/aas:2000147},
       adsurl = {https://ui.adsabs.harvard.edu/abs/2000A&AS..142..181C}
}

@ARTICLE{yang2023,
       author = {{Yang}, Chentao and {Omont}, Alain and {Mart{\'\i}n}, Sergio and {Bisbas}, Thomas G. and {Cox}, Pierre and {Beelen}, Alexandre and {Gonz{\'a}lez-Alfonso}, Eduardo and {Gavazzi}, Rapha{\"e}l and {Aalto}, Susanne and {Andreani}, Paola and {Ceccarelli}, Cecilia and {Gao}, Yu and {Gorski}, Mark and {Gu{\'e}lin}, Michel and {Fu}, Hai and {Ivison}, R.~J. and {Knudsen}, Kirsten K. and {Lehnert}, Matthew and {Messias}, Hugo and {Muller}, Sebastien and {Neri}, Roberto and {Riechers}, Dominik and {van der Werf}, Paul and {Zhang}, Zhi-Yu},
        title = "{SUNRISE: The rich molecular inventory of high-redshift dusty galaxies revealed by broadband spectral line surveys}",
      journal = {\aap},
         year = 2023,
        month = dec,
       volume = {680},
          eid = {A95},
        pages = {A95},
          doi = {10.1051/0004-6361/202347610},
archivePrefix = {arXiv},
       eprint = {2308.07368},
 primaryClass = {astro-ph.GA},
       adsurl = {https://ui.adsabs.harvard.edu/abs/2023A&A...680A..95Y}
}

@ARTICLE{Sewilo2022,
       author = {{Sewi{\l}o}, Marta and {Cordiner}, Martin and {Charnley}, Steven B. and {Oliveira}, Joana M. and {Garcia-Berrios}, Emmanuel and {Schilke}, Peter and {Ward}, Jacob L. and {Wiseman}, Jennifer and {Indebetouw}, Remy and {Tokuda}, Kazuki and {van Loon}, Jacco Th. and {S{\'a}nchez-Monge}, {\'A}lvaro and {Allen}, Veronica and {Chen}, C. -H. Rosie and {Hamedani Golshan}, Roya and {Karska}, Agata and {Kristensen}, Lars E. and {Kurtz}, Stan E. and {M{\"o}ller}, Thomas and {Onishi}, Toshikazu and {Zahorecz}, Sarolta},
        title = "{ALMA Observations of Molecular Complexity in the Large Magellanic Cloud: The N 105 Star-forming Region}",
      journal = {\apj},
         year = 2022,
        month = jun,
       volume = {931},
       number = {2},
          eid = {102},
        pages = {102},
          doi = {10.3847/1538-4357/ac4e8f},
archivePrefix = {arXiv},
       eprint = {2201.09945},
 primaryClass = {astro-ph.GA},
       adsurl = {https://ui.adsabs.harvard.edu/abs/2022ApJ...931..102S}
}

@ARTICLE{maret2005,
       author = {{Maret}, S. and {Ceccarelli}, C. and {Tielens}, A.~G.~G.~M. and {Caux}, E. and {Lefloch}, B. and {Faure}, A. and {Castets}, A. and {Flower}, D.~R.},
        title = "{CH3OH abundance in low mass protostars}",
      journal = {\aap},
         year = 2005,
        month = nov,
       volume = {442},
       number = {2},
        pages = {527-538},
          doi = {10.1051/0004-6361:20052899},
archivePrefix = {arXiv},
       eprint = {astro-ph/0507172},
 primaryClass = {astro-ph},
       adsurl = {https://ui.adsabs.harvard.edu/abs/2005A&A...442..527M}
}

@ARTICLE{gerin2025,
       author = {{Gerin}, Maryvonne and {Liszt}, Harvey and {Tercero}, Bel{\'e}n and {Cernicharo}, Jos{\'e}},
        title = "{Spectral survey of the diffuse gas toward BL Lac in the Q band}",
      journal = {\aap},
         year = 2025,
        month = nov,
       volume = {703},
          eid = {A125},
        pages = {A125},
          doi = {10.1051/0004-6361/202556101},
archivePrefix = {arXiv},
       eprint = {2509.14822},
 primaryClass = {astro-ph.GA},
       adsurl = {https://ui.adsabs.harvard.edu/abs/2025A&A...703A.125G}
}

@ARTICLE{bianchi2019,
       author = {{Bianchi}, E. and {Codella}, C. and {Ceccarelli}, C. and {Vazart}, F. and {Bachiller}, R. and {Balucani}, N. and {Bouvier}, M. and {De Simone}, M. and {Enrique-Romero}, J. and {Kahane}, C. and {Lefloch}, B. and {L{\'o}pez-Sepulcre}, A. and {Ospina-Zamudio}, J. and {Podio}, L. and {Taquet}, V.},
        title = "{The census of interstellar complex organic molecules in the Class I hot corino of SVS13-A}",
      journal = {\mnras},
         year = 2019,
        month = feb,
       volume = {483},
       number = {2},
        pages = {1850-1861},
          doi = {10.1093/mnras/sty2915},
archivePrefix = {arXiv},
       eprint = {1810.11411},
 primaryClass = {astro-ph.SR},
       adsurl = {https://ui.adsabs.harvard.edu/abs/2019MNRAS.483.1850B}
}

@ARTICLE{madhusudhan2019,
       author = {{Madhusudhan}, Nikku},
        title = "{Exoplanetary Atmospheres: Key Insights, Challenges, and Prospects}",
      journal = {\araa},
         year = 2019,
        month = aug,
       volume = {57},
        pages = {617-663},
          doi = {10.1146/annurev-astro-081817-051846},
archivePrefix = {arXiv},
       eprint = {1904.03190},
 primaryClass = {astro-ph.EP},
       adsurl = {https://ui.adsabs.harvard.edu/abs/2019ARA&A..57..617M}
}

@ARTICLE{sanz-novo2026,
       author = {{Sanz-Novo}, M. and {Rivilla}, V.~M. and {Jim{\'e}nez-Serra}, I. and {Colzi}, L. and {Zeng}, S. and {Meg{\'\i}as}, A. and {San Andr{\'e}s}, D. and {L{\'o}pez-Gallifa}, {\'A}. and {Mart{\'\i}nez-Henares}, A. and {Fried}, Z.~T.~P. and {McGuire}, B.~A. and {Mart{\'\i}n}, S. and {Requena-Torres}, M.~A. and {Tercero}, B. and {de Vicente}, P. and {Kolesnikov{\'a}}, L. and {Alonso}, E.~R. and {Cocinero}, E.~J. and {Guillemin}, J.~C. and {Kleiner}, I.},
        title = "{Expanding the C$_{3}$H$_{6}$O$_{2}$ isomeric interstellar inventory: Discovery of lactaldehyde and methoxyacetaldehyde in G+0.693-0.027}",
      journal = {\aap},
         year = 2026,
        month = feb,
       volume = {706},
          eid = {A316},
        pages = {A316},
          doi = {10.1051/0004-6361/202558316},
archivePrefix = {arXiv},
       eprint = {2601.07365},
 primaryClass = {astro-ph.GA},
       adsurl = {https://ui.adsabs.harvard.edu/abs/2026A&A...706A.316S}
}

@ARTICLE{matsuura2017,
       author = {{Matsuura}, M. and {Indebetouw}, R. and {Woosley}, S. and {Bujarrabal}, V. and {Abell{\'a}n}, F.~J. and {McCray}, R. and {Kamenetzky}, J. and {Fransson}, C. and {Barlow}, M.~J. and {Gomez}, H.~L. and {Cigan}, P. and {De Looze}, I. and {Spyromilio}, J. and {Staveley-Smith}, L. and {Zanardo}, G. and {Roche}, P. and {Larsson}, J. and {Viti}, S. and {van Loon}, J. Th. and {Wheeler}, J.~C. and {Baes}, M. and {Chevalier}, R. and {Lundqvist}, P. and {Marcaide}, J.~M. and {Dwek}, E. and {Meixner}, M. and {Ng}, C.-Y. and {Sonneborn}, G. and {Yates}, J.},
        title = "{ALMA spectral survey of Supernova 1987A - molecular inventory, chemistry, dynamics and explosive nucleosynthesis}",
      journal = {\mnras},
         year = 2017,
        month = aug,
       volume = {469},
       number = {3},
        pages = {3347-3362},
          doi = {10.1093/mnras/stx830},
archivePrefix = {arXiv},
       eprint = {1704.02324},
 primaryClass = {astro-ph.SR},
       adsurl = {https://ui.adsabs.harvard.edu/abs/2017MNRAS.469.3347M}
}

@ARTICLE{fayolle2017,
       author = {{Fayolle}, Edith C. and {{\"O}berg}, Karin I. and {J{\o}rgensen}, Jes K. and {Altwegg}, Kathrin and {Calcutt}, Hannah and {M{\"u}ller}, Holger S.~P. and {Rubin}, Martin and {van der Wiel}, Matthijs H.~D. and {Bjerkeli}, Per and {Bourke}, Tyler L. and {Coutens}, Audrey and {van Dishoeck}, Ewine F. and {Drozdovskaya}, Maria N. and {Garrod}, Robin T. and {Ligterink}, Niels F.~W. and {Persson}, Magnus V. and {Wampfler}, Susanne F. and {Rosina Team}},
        title = "{Protostellar and cometary detections of organohalogens}",
      journal = {Nature Astronomy},
         year = 2017,
        month = oct,
       volume = {1},
        pages = {703-708},
          doi = {10.1038/s41550-017-0237-7},
       adsurl = {https://ui.adsabs.harvard.edu/abs/2017NatAs...1..703F}
}

@ARTICLE{mcguire2017,
       author = {{McGuire}, Brett A. and {Shingledecker}, Christopher N. and {Willis}, Eric R. and {Burkhardt}, Andrew M. and {El-Abd}, Samer and {Motiyenko}, Roman A. and {Brogan}, Crystal L. and {Hunter}, Todd R. and {Margul{\`e}s}, Laurent and {Guillemin}, Jean-Claude and {Garrod}, Robin T. and {Herbst}, Eric and {Remijan}, Anthony J.},
        title = "{ALMA Detection of Interstellar Methoxymethanol (CH$_{3}$OCH$_{2}$OH)}",
      journal = {\apjl},
         year = 2017,
        month = dec,
       volume = {851},
       number = {2},
          eid = {L46},
        pages = {L46},
          doi = {10.3847/2041-8213/aaa0c3},
archivePrefix = {arXiv},
       eprint = {1712.03256},
 primaryClass = {astro-ph.GA},
       adsurl = {https://ui.adsabs.harvard.edu/abs/2017ApJ...851L..46M}
}

@ARTICLE{coutens2019,
       author = {{Coutens}, A. and {Ligterink}, N.~F.~W. and {Loison}, J.-C. and {Wakelam}, V. and {Calcutt}, H. and {Drozdovskaya}, M.~N. and {J{\o}rgensen}, J.~K. and {M{\"u}ller}, H.~S.~P. and {van Dishoeck}, E.~F. and {Wampfler}, S.~F.},
        title = "{The ALMA-PILS survey: First detection of nitrous acid (HONO) in the interstellar medium}",
      journal = {\aap},
         year = 2019,
        month = mar,
       volume = {623},
          eid = {L13},
        pages = {L13},
          doi = {10.1051/0004-6361/201935040},
archivePrefix = {arXiv},
       eprint = {1903.03378},
 primaryClass = {astro-ph.SR},
       adsurl = {https://ui.adsabs.harvard.edu/abs/2019A&A...623L..13C}
}

@ARTICLE{yan2020,
       author = {{Zhou}, Yan and {Quan}, Dong-Hui and {Zhang}, Xia and {Qin}, Sheng-Li},
        title = "{Detection of hydroxyacetone in protostar IRAS 16293-2422 B}",
      journal = {Research in Astronomy and Astrophysics},
         year = 2020,
        month = aug,
       volume = {20},
       number = {8},
          eid = {125},
        pages = {125},
          doi = {10.1088/1674-4527/20/8/125},
       adsurl = {https://ui.adsabs.harvard.edu/abs/2020RAA....20..125Z}
}

@ARTICLE{coutens2022,
       author = {{Coutens}, A. and {Loison}, J.-C. and {Boulanger}, A. and {Caux}, E. and {M{\"u}ller}, H.~S.~P. and {Wakelam}, V. and {Manigand}, S. and {J{\o}rgensen}, J.~K.},
        title = "{The ALMA-PILS survey: First tentative detection of 3-hydroxypropenal (HOCHCHCHO) in the interstellar medium and chemical modeling of the C$_{3}$H$_{4}$O$_{2}$ isomers}",
      journal = {\aap},
         year = 2022,
        month = apr,
       volume = {660},
          eid = {L6},
        pages = {L6},
          doi = {10.1051/0004-6361/202243038},
archivePrefix = {arXiv},
       eprint = {2203.14119},
 primaryClass = {astro-ph.GA},
       adsurl = {https://ui.adsabs.harvard.edu/abs/2022A&A...660L...6C}
}

@ARTICLE{muller2024,
       author = {{Muller}, S. and {Le Gal}, R. and {Roueff}, E. and {Black}, J.~H. and {Faure}, A. and {Gu{\'e}lin}, M. and {Omont}, A. and {G{\'e}rin}, M. and {Combes}, F. and {Aalto}, S.},
        title = "{Protonated acetylene in the z = 0.89 molecular absorber toward PKS 1830-211}",
      journal = {\aap},
         year = 2024,
        month = mar,
       volume = {683},
          eid = {A62},
        pages = {A62},
          doi = {10.1051/0004-6361/202348994},
archivePrefix = {arXiv},
       eprint = {2401.09975},
 primaryClass = {astro-ph.GA},
       adsurl = {https://ui.adsabs.harvard.edu/abs/2024A&A...683A..62M}
}

@ARTICLE{Tasa-Chaveli2025,
       author = {{Tasa-Chaveli}, A. and {S{\'a}nchez-Monge}, {\'A}. and {Fuente}, A. and {Ginsburg}, A. and {M{\"u}ller}, H.~S.~P. and {M{\"o}ller}, T. and {Rivi{\`e}re-Marichalar}, P. and {Navarro-Almaida}, D. and {Esplugues}, G. and {Schilke}, P. and {Rodr{\'\i}guez-Baras}, M. and {Thorwirth}, S. and {Beitia-Antero}, L.},
        title = "{A Quest for Sulfur-bearing Refractory Species: Identification of CaS in the Interstellar Medium}",
      journal = {\apjl},
         year = 2025,
        month = nov,
       volume = {993},
       number = {2},
          eid = {L42},
        pages = {L42},
          doi = {10.3847/2041-8213/ae0e6b},
archivePrefix = {arXiv},
       eprint = {2510.02937},
 primaryClass = {astro-ph.GA},
       adsurl = {https://ui.adsabs.harvard.edu/abs/2025ApJ...993L..42T}
}

@ARTICLE{duan2026,
       author = {{Duan}, Chunguo and {Xu}, Fengwei and {Kang}, Jun and {Gou}, Qian and {Xu}, Xuefang and {Pagani}, Laurent and {Du}, Jiaxin and {Chen}, Xi},
        title = "{First Interstellar Detection of Methyl Carbamate: A New Observational Anchor for Glycine Chemistry}",
      journal = {\apjl},
         year = 2026,
        month = jun,
       volume = {1004},
       number = {1},
          eid = {L6},
        pages = {L6},
          doi = {10.3847/2041-8213/ae69da},
archivePrefix = {arXiv},
       eprint = {2605.07159},
 primaryClass = {astro-ph.GA},
       adsurl = {https://ui.adsabs.harvard.edu/abs/2026ApJ..1004L...6D}
}

@ARTICLE{zhang2026,
       author = {{Zhang}, Ziwei E. and {Sakai}, Nami and {Agundez}, Marcelino and {Podio}, Linda and {Codella}, Claudio and {Hirota}, Tomoya and {Ginsburg}, Adam and {McGuire}, Brett and {Murillo}, Nadia M. and {Okoda}, Yuki and {Oya}, Yoko and {Sabatini}, Giovanni and {Takigawa}, Aki and {Tanaka}, Kei E.~I. and {Zhang}, Yichen and {Luo}, Qiuyi and {Cheng}, Yu and {Jimenez-Serra}, Izaskun and {Tachibana}, Shogo},
        title = "{Dust Seeding Molecules in a Massive Protostar -- Detection of TiO in Orion Source I}",
      journal = {arXiv e-prints},
         year = 2026,
        month = aug,
          eid = {arXiv:2608.01505},
        pages = {arXiv:2608.01505},
          doi = {10.48550/arXiv.2608.01505},
archivePrefix = {arXiv},
       eprint = {2608.01505},
 primaryClass = {astro-ph.GA},
       adsurl = {https://ui.adsabs.harvard.edu/abs/2026arXiv260801505Z}
}

@ARTICLE{tachibana2019,
       author = {{Tachibana}, Shogo and {Kamizuka}, Takafumi and {Hirota}, Tomoya and {Sakai}, Nami and {Oya}, Yoko and {Takigawa}, Aki and {Yamamoto}, Satoshi},
        title = "{Spatial Distribution of AlO in a High-mass Protostar Candidate Orion Source I}",
      journal = {\apjl},
         year = 2019,
        month = apr,
       volume = {875},
       number = {2},
          eid = {L29},
        pages = {L29},
          doi = {10.3847/2041-8213/ab1653},
       adsurl = {https://ui.adsabs.harvard.edu/abs/2019ApJ...875L..29T}
}

@article{cabezas2012,
    author = {{Cabezas}, Carlos and Varela, Marcelino and Peña, Isabel and Mata, Santiago and López, Juan C. and Alonso, José L.},
    title = {The conformational locking of asparagine},
    journal = {Chemical Communications},
    volume = {48},
    number = {47},
    pages = {5934-5936},
    year = {2012},
    month = {06},
    issn = {1359-7345},
    doi = {10.1039/c2cc32109b},
    url = {https://doi.org/10.1039/c2cc32109b},
    eprint = {https://pubs.rsc.org/cc/article-pdf/48/47/5934/2534212/c2cc32109b.pdf},
}

@article{pearce2018,
author = {{Pearce}, Ben K.D. and Tupper, Andrew S. and Pudritz, Ralph E. and Higgs, Paul G.},
title = {Constraining the Time Interval for the Origin of Life on Earth},
journal = {Astrobiology},
volume = {18},
number = {3},
pages = {343-364},
year = {2018},
doi = {10.1089/ast.2017.1674},
    note ={PMID: 29570409},

URL = { 
        https://doi.org/10.1089/ast.2017.1674
    
},
eprint = { 
        https://doi.org/10.1089/ast.2017.1674
    
}
}

@ARTICLE{lodders2003,
       author = {{Lodders}, Katharina},
        title = "{Solar System Abundances and Condensation Temperatures of the Elements}",
      journal = {\apj},
         year = 2003,
        month = jul,
       volume = {591},
       number = {2},
        pages = {1220-1247},
          doi = {10.1086/375492},
       adsurl = {https://ui.adsabs.harvard.edu/abs/2003ApJ...591.1220L}
}

@ARTICLE{berne2023,
       author = {{Bern{\'e}}, Olivier and {Martin-Drumel}, Marie-Aline and {Schroetter}, Ilane and {Goicoechea}, Javier R. and {Jacovella}, Ugo and {Gans}, B{\'e}renger and {Dartois}, Emmanuel and {Coudert}, Laurent H. and {Bergin}, Edwin and {Alarcon}, Felipe and {Cami}, Jan and {Roueff}, Evelyne and {Black}, John H. and {Asvany}, Oskar and {Habart}, Emilie and {Peeters}, Els and {Canin}, Amelie and {Trahin}, Boris and {Joblin}, Christine and {Schlemmer}, Stephan and {Thorwirth}, Sven and {Cernicharo}, Jose and {Gerin}, Maryvonne and {Tielens}, Alexander and {Zannese}, Marion and {Abergel}, Alain and {Bernard-Salas}, Jeronimo and {Boersma}, Christiaan and {Bron}, Emeric and {Chown}, Ryan and {Cuadrado}, Sara and {Dicken}, Daniel and {Elyajouri}, Meriem and {Fuente}, Asunci{\'o}n and {Gordon}, Karl D. and {Issa}, Lina and {Kannavou}, Olga and {Khan}, Baria and {Lacinbala}, Ozan and {Languignon}, David and {Le Gal}, Romane and {Maragkoudakis}, Alexandros and {Meshaka}, Raphael and {Okada}, Yoko and {Onaka}, Takashi and {Pasquini}, Sofia and {Pound}, Marc W. and {Robberto}, Massimo and {R{\"o}llig}, Markus and {Schefter}, Bethany and {Schirmer}, Thi{\'e}baut and {Sidhu}, Ameek and {Tabone}, Benoit and {Van De Putte}, Dries and {Vicente}, S{\'\i}lvia and {Wolfire}, Mark G.},
        title = "{Formation of the methyl cation by photochemistry in a protoplanetary disk}",
      journal = {\nat},
         year = 2023,
        month = sep,
       volume = {621},
       number = {7977},
        pages = {56-59},
          doi = {10.1038/s41586-023-06307-x},
archivePrefix = {arXiv},
       eprint = {2401.03296},
 primaryClass = {astro-ph.GA},
       adsurl = {https://ui.adsabs.harvard.edu/abs/2023Natur.621...56B}
}

@ARTICLE{dunham1937,
       author = {{Dunham}, Jr., T.},
        title = "{Interstellar Neutral Potassium and Neutral Calcium}",
      journal = {\pasp},
         year = 1937,
        month = feb,
       volume = {49},
       number = {287},
        pages = {26-28},
          doi = {10.1086/124759},
       adsurl = {https://ui.adsabs.harvard.edu/abs/1937PASP...49...26D}
}

\end{document}